\documentclass[iop,preprint2]{emulateapj1}
\usepackage{psfig}
\usepackage{apjfonts}
\usepackage{ulem}
\usepackage[dvips]{color}
\usepackage{graphicx}
\usepackage{rotating,tabularx}
\usepackage[USenglish]{babel}
\usepackage{mathrsfs}
\usepackage{longtable}

\newcommand{\masyr}{${\rm mas}\,{\rm yr}^{-1}$}

\newcommand{\wcen}{$\omega$Cen}

\begin{document}

\title{THE STATE-OF-THE-ART \textit{HST} ASTRO-PHOTOMETRIC ANALYSIS OF
  THE CORE OF $\omega$~CENTAURI. I. THE CATALOG$^{\ast}$}

\author{
A.\ Bellini\altaffilmark{1},
J.\ Anderson\altaffilmark{1}, 
L.\ R.\ Bedin\altaffilmark{2},
I.\ R.\ King\altaffilmark{3},
R.\ P.\ van der Marel\altaffilmark{1},
G.\ Piotto\altaffilmark{2,4},
and A.\ Cool\altaffilmark{5}}

\altaffiltext{1}{Space Telescope Science Institute, 3700 San Martin
  Dr., Baltimore, MD 21218, USA} \altaffiltext{2}{Istituto Nazionale
  di Astrofisica, Osservatorio Astronomico di Padova, v.co
  dell'Osservatorio 5, Padova, I-35122, Italy}
\altaffiltext{3}{Department of Astronomy, University of Washington,
  Box 351580, Seattle, 98195, WA, USA} \altaffiltext{4}{Dipartimento
  di Fisica e Astronomia ``Galileo Galilei'', Universit\`a di Padova,
  Vicolo dell'Osservatorio 3, Padova I-35122, Italy}
\altaffiltext{5}{Department of Physics and Astronomy, San Francisco
  State University, 1600 Holloway Ave., San Francisco, CA 94132, USA}
\altaffiltext{$^\ast$}{Based on archival observations with the
  NASA/ESA \textit{Hubble Space Telescope}, obtained at the Space
  Telescope Science Institute, which is operated by AURA, Inc., under
  NASA contract NAS 5-26555.}

\received{October 8, 2016}
\revised{April 7, 2017}
\accepted{April 11, 2017}

\email{bellini@stsci.edu}

\begin{abstract}
We have constructed the most-comprehensive catalog of photometry and
proper motions ever assembled for a globular cluster (GC). The core of
\wcen\ has been imaged over 650 times through WFC3's UVIS and IR
channels for the purpose of detector calibration. There exist from 4
to over 60 exposures through each of 26 filters, stretching
continuously from F225W in the UV to F160W in the
infrared. Furthermore, the 11-year baseline between these data and a
2002 ACS survey has allowed us to more than double the proper-motion
accuracy and triple the number of well-measured stars compared to our
previous groundbreaking effort. This totally unprecedented complete
spectral coverage for over 470$\,$000 stars within the cluster's core,
from the tip of the red-giant branch down to the white dwarfs,
provides the best astro-photometric observational data base yet to
understand the multiple-population phenomenon in any GC. In this first
paper of the series we describe in detail the data-reduction processes
and deliver the astro-photometric catalog to the astronomical
community.\\

\end{abstract}


\maketitle

\begin{table*}[!th]
\label{tab:log}
\centering
{
\begin{tabular}{lcllcc}
\multicolumn{6}{c}{\textsc{Table~1}}\\
\multicolumn{6}{c}{\textsc{List of \textit{HST} WFC3 Observations of the Core of
    NGC 5139}}\\
\hline\hline
\textbf{Filter}&\textbf{Exposures}&\textbf{Total Time}&\textbf{Program
  ID}&\textbf{Description}&\textbf{Epoch}\\
\hline
\multicolumn{6}{c}{WFC3/UVIS}\\
\hline
F225W &9$\times$350$\,$s + 27$\times$900$\,$s&27$\,$450$\,$s&
11452, 11911, 12339& UV Wide & 2009--2011\\
F275W &9$\times$350$\,$s + 31$\times$800$\,$s&27$\,$600$\,$s&
11452, 11911, 12339& UV Wide   & 2009--2011\\
F336W &11$\times$10$\,$s + 37$\times$350$\,$s&13$\,$060$\,$s&
11452, 11911, 12339, 12802&  $U$, Str\"{o}mgren $u$ & 2009--2011\\
F350LP&4$\times$350$\,$s&1400$\,$s& 12353 &Long pass& 2010--2011\\
F390M &4$\times$350$\,$s&1400$\,$s& 12353 &Ca\textsc{ii} continuum & 2010--2011\\
F390W &15$\times$350$\,$s&5250$\,$s& 11911 &Washington $C$ & 2010\\
F438W &34$\times$350$\,$s&11$\,$900$\,$s& 11911, 12339 &Johnson $B$& 2010--2011\\
F467M &3$\times$400$\,$s + 3$\times$450$\,$s&2250$\,$s&12694&
Str\"{o}mgren $b$& 2012\\
F555W &27$\times$40$\,$s&960$\,$s& 11911, 12339 &Johnson $V$& 2010--2011\\
F606W &1$\times$35$\,$s + 55$\times$40$\,$s &2523$\,$s&11452, 11911,
12094, 12339,&Wide Johnson $V$& 2009--2013\\
      &+ 6$\times$48$\,$s                   &         &12353, 12694, 12714, 13100&&\\
F621M &6$\times$708$\,$s&4248$\,$s& 13100 &11\% passband& 2012--2013\\
F656N &5$\times$500$\,$s&2500$\,$s& 11922 &H$\alpha$@\AA 6562& 2010\\
F658N &6$\times$350$\,$s&1400$\,$s& 12091 &N\textsc{ii}@\AA 6583& 2010\\
F673N &6$\times$350$\,$s + 3$\times$400$\,$s&2600$\,$s& 12091, 12694
&S\textsc{ii}@\AA 6717/6731& 2010--2012\\
F775W &16$\times$350$\,$s + 2$\times$450$\,$s&6500$\,$s& 11911, 12700&SDSS $i'$& 2010--2012\\
F814W &1$\times$35$\,$s + 33$\times$40$\,$s&1355$\,$s& 11452, 11911,
12339 &Wide Johnson $I$& 2009--2011\\
F850LP&26$\times$60$\,$s&1560$\,$s& 11911, 12339 &SDSS $z'$& 2010--2011\\
F953N &5$\times$850$\,$s&4250$\,$s& 11922 &S\textsc{iii}@\AA 9532& 2010\\
\hline
\multicolumn{6}{c}{WFC3/IR}\\
\hline
F098M&40$\times$352$\,$s + 4$\times$602$\,$s&16$\,$488$\,$s& 11928, 12340 &Blue grism ref.& 2009--2012\\
F105W&4$\times$227$\,$s&908$\,$s& 12353 &Wide $Y$& 2010--2011\\
F110W&41$\times$227$\,$s + 3$\times$274$\,$s&10$\,$129$\,$s& 11928,
12340, 12351 &Wide $YJ$& 2010--2012\\
F125W&41$\times$227$\,$s + 30$\times$349$\,$s&19$\,$777$\,$s& 11928,
12340, 12694 &Wide $J$& 2010--2012\\
F139M&40$\times$502$\,$s&20$\,$080$\,$s& 11928, 12340 &H$_2$O/CH$_4$ line& 2010--2012\\
F140W&4$\times$502$\,$s&2008$\,$s& 12353 &$\!\!\!\!\!\!\!\!\!\!$Wide $JH$ gap, Red
grism ref.& 2010--2012\\
F153M&4$\times$252$\,$s&1008$\,$s& 12353 &H$_2$O 
and NH$_3$& 2010--2012\\
F160W&59$\times$252$\,$s&14$\,$868$\,$s& 11928, 12340, 12353, 12714 &$H$& 2010--2012\\
\hline\hline
\end{tabular}}
\end{table*}

\begin{figure*}[!t]
\centering
\includegraphics[width=17cm]{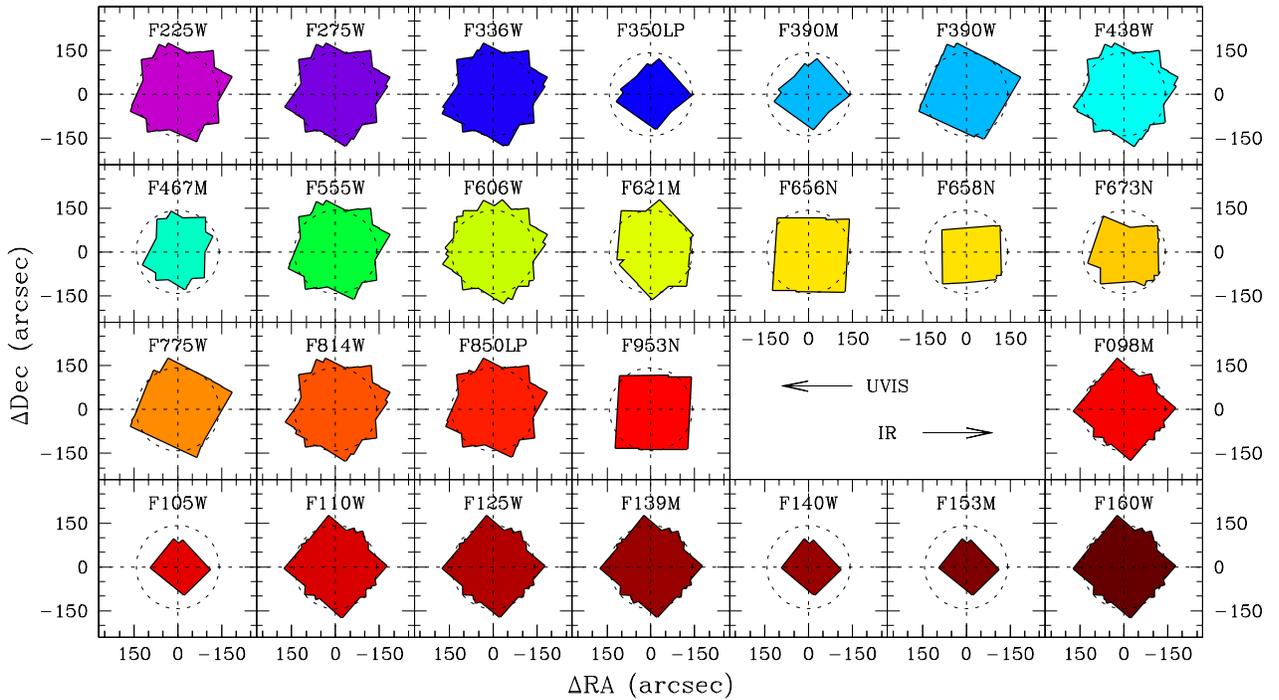}
\caption{\small{Total FoVs covered by each WFC3 filter, in order of
    increasing central wavelength. The footprints are color-coded
    according to the filters' central wavelength, from violet (F225W)
    to green (F555W) to brown (F160W). Axes are in arcsecs with
    respect to the cluster's center. The dotted circle in each panel
    highlights the cluster's core radius ($r_{\rm c}=2\farcm37$,
    \citealt{1996AJ....112.1487H}, 2010 edition).}}
\label{f:fovs}
\end{figure*}

\begin{figure*}[!t]
\centering
\includegraphics[width=16.5cm]{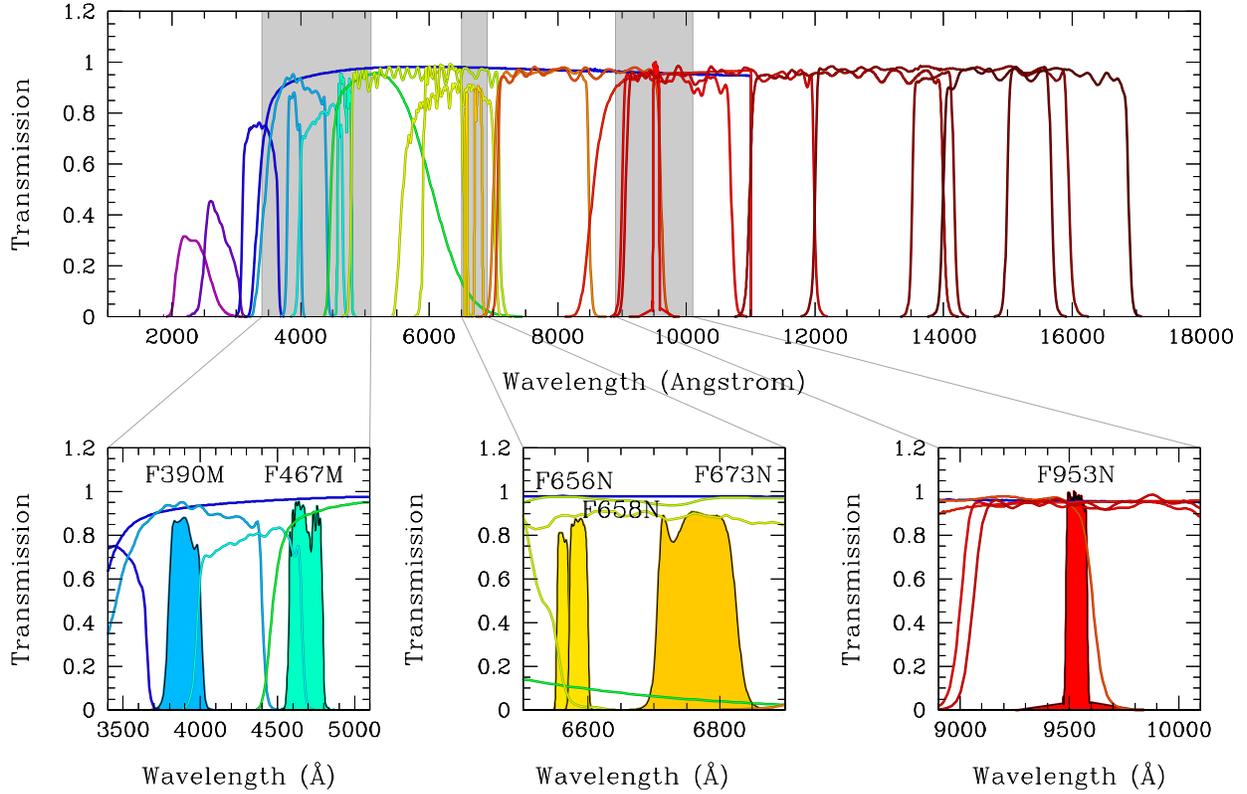}
\caption{\small{Transmission curves of the 26 filters, color coded as
    in Fig.~\ref{f:fovs}, providing a complete wavelength coverage
    from 200 to 1750 nm. The three bottom panels offer a zoomed-in
    view around the transmission curves of medium- and narrow-band
    filters of the UVIS camera.}}
\label{f:tc}
\end{figure*}

\begin{figure*}[!t]
\centering
\includegraphics[width=\textwidth]{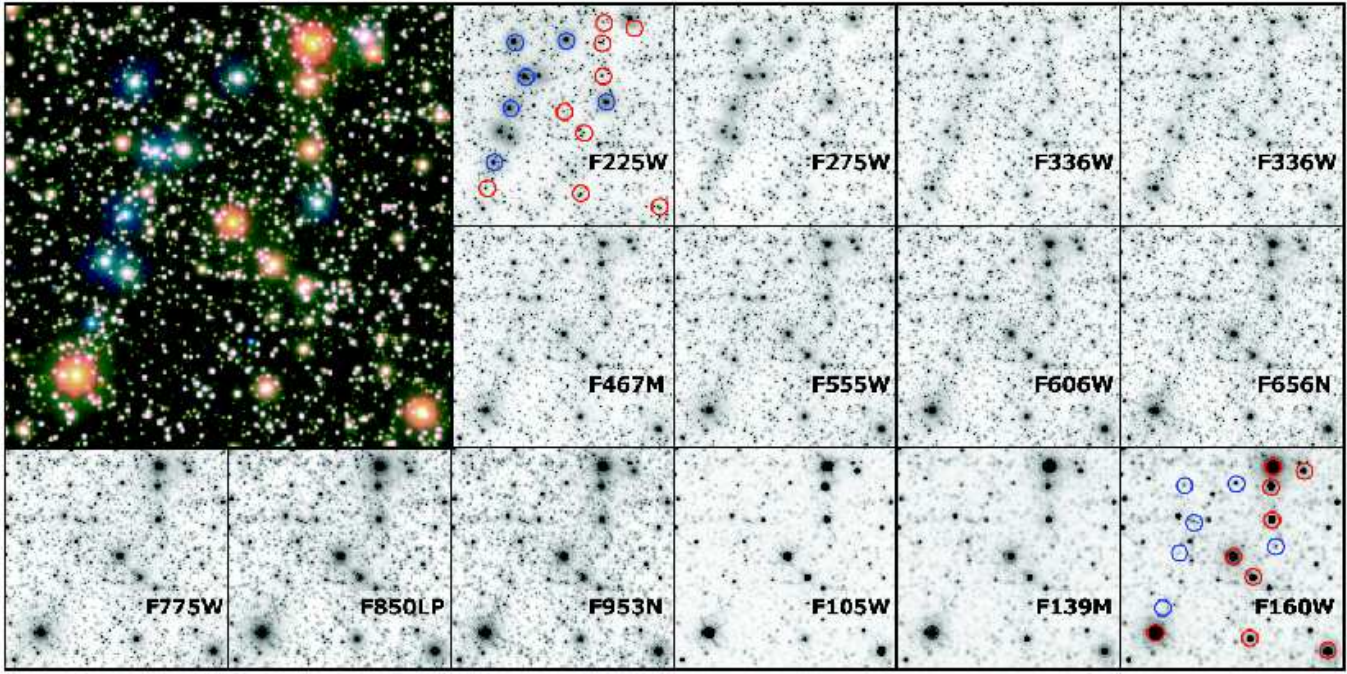}
\caption{\small{The top-left panel shows a trichromatic (Red=F814W,
    Green=F606W, Blue=F275W)
    $20^{\prime\prime}$$\times$$20^{\prime\prime}$ region of the FoV
    close to the center of the cluster. This region contains both HB
    (bright blue) and RGB (bright red) stars.  The remaining panels
    show the same region as seen through a selection of different
    filters. HB stars (marked with blue circles in the first and the
    last of these panels) become dimmer and dimmer moving toward
    redder filters. On the other hand, RGB stars (marked in red)
    become increasingly bright.}}
\label{f:tiles}
\end{figure*}

\begin{figure*}[!t]
\centering
\includegraphics[width=16cm]{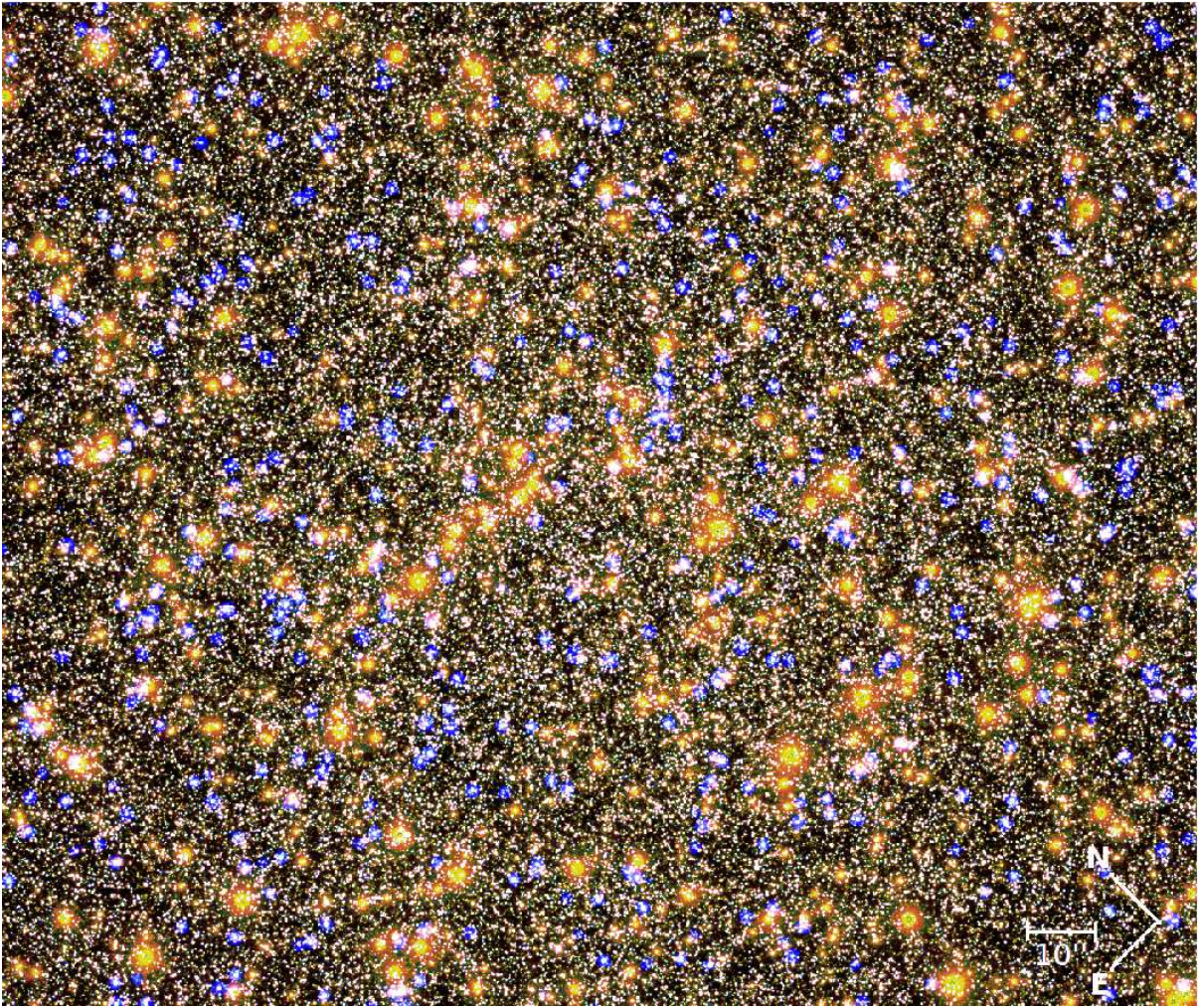}
\caption{\small{Trichromatic stack of the inner
    3$^\prime$$\times$$2\farcm5$ region of \wcen, made using the
    monochromatic stacks of bluest, the reddest, and an intermediate
    filter (blue=F225W, green=F814W, red=F160W). Scale
    (10$^{\prime\prime}$) and orientation are shown on the
    bottom-right corner.}}
\label{f:trichro}
\end{figure*}

\section{Introduction}
\label{sec:1}

Based on studies of the color-magnitude diagram (CMD), a decade ago
NGC 5139 (\wcen) was the only Galactic globular cluster (GC) known to
contain multiple populations (\citealt{anderson97,
  1999Natur.402...55L, 2000ApJ...534L..83P,
  2004ApJ...605L.125B}).\footnote[1]{From the spectroscopic point of
  view, \wcen\ was known to host stars with a wide range in
  metallicity since at least the 70s (e.g.,
  \citealt{1973MNRAS.162..207C, 1975ApJ...201L..75N,
    1975ApJ...201L..71F}).}  At the time, it was not clear whether it
represented a transition object between clusters and galaxies or
simply a merger of two clusters (\citealt{2002ApJ...573L..95F,
  2003ApJ...591L.127P, 2006ApJ...637L.109B}). We now know that
\wcen\ is not a solitary exception to the single-population rule but
rather is just an extreme example of a self-enrichment phenomenon that
is present at some level in all globular clusters (e.g.,
\citealt{2015AJ....149...91P}, and references therein).  The
\textit{Hubble Space Telescope} (\textit{HST}) has played a pivotal
role both in helping to clarify the confusing situation in \wcen\ and
in allowing us to look for the presence of distinct sequences and
intrinsic spreads in all evolutionary phases.

While a detailed study of \wcen\ has pointed the way to identifying
multiple stellar populations (MPs) in other clusters, even with so
many other MP examples we still do not know how they formed, either in
\wcen\ or in other clusters (\citealt{2015MNRAS.454.4197R}). Once
again, \wcen\ may represent our best laboratory to shed light on the
phenomenon. It is nearby, and is one of the few GCs that have not had
time to fully relax (\citealt{2007ApJ...654..915S,
  2009A&A...493..959B, 2009A&A...507.1393B, 2010ApJ...710.1032A}), so
that the motions of its stars still retain some memory of their origin
(see, e.g., \citealt{2013ApJ...771L..15R, 2015ApJ...810L..13B}).  By
studying the spatial distributions, rotations, and anisotropies of
\wcen's populations, we can reconstruct its star-formation history in
a way that is not possible for other clusters.

The single greatest boon to our understanding of \wcen\ has come from
program GO-9442 (PI:\ A.\ Cool), which imaged the central
$600^{\prime\prime}$$\times$$600^{\prime\prime}$ (about 2 core radii)
with the Wide-Field Channel of the Advanced Camera for Surveys
(ACS/WFC) in F435W (\textit{B}), F625W (\textit{R}), and F658N
(H$\alpha$), soon after ACS was installed during \textit{HST}'s
Service Mission 3B. This large data set, which contains more than two
million stars, has resulted in over 25 publications by over ten
different teams and has contributed immeasurably to our understanding
of the main sequence (MS), sub-giant branch (SGB), red-giant branch
(RGB), horizontal branch (HB), and white dwarf (WD) populations in
this enigmatic cluster. Some highlights include: (1) the existence of
the anomalous lower turnoff (\citealt{2004ApJ...603L..81F}); (2) the
splitting of the MS (\citealt{2004ApJ...605L.125B}); (3) optical
counterparts for X-ray sources (\citealt{2004ApJ...613..512H}); (4)
evidence for prolonged star formation (\citealt{2007ApJ...663..296V});
(5) radial gradients among MS populations
(\citealt{2009A&A...507.1393B}); (6) investigating the presence of a
central intermediate-mass black hole (\citealt{2010ApJ...710.1032A,
  2010ApJ...710.1063V}); and (7) analysis of the blue-hook stars in
the HB (\citealt{2015Natur.523..318T}).

Notwithstanding these many multi-faceted investigations, there is a
clear evidence that \wcen\ has not yet disclosed all of its
secrets. Most of the progress thus far has come largely from a single
set of WFC/ACS images optimized to identify cataclysmic variables
(CVs).  As it turnes out, the calibration team of the newly installed
Wide-Field Camera 3 (WFC3) has decided to use the center of this
cluster as one of its fundamental calibration targets. \wcen\ is the
only cluster for which the core diameter is larger than the ACS or
WFC3 field of view (FoV), making it a uniquely excellent target for
distortion and flat-field calibration, particularly given its
``Goldilocks'' star density.  The plentiful UV-bright stars on the
cluster's hot-HB provide a large number of sources to calibrate the
bluest filters (see, e.g. \citealt{2009PASP..121.1419B}).

Since the installation of WFC3 in Service Mission 4, its two
Ultraviolet-VISible (UVIS) and InfraRed (IR) channels have observed
the center of \wcen\ in over 30 different visits\footnote[2]{Up to mid
  2013, see Table~1.}, generating more than 650 individual exposures.
In \cite{2011PASP..123..622B}, we used some of this data to model the
three sources of distortion (camera optics, filter-specific residuals,
and a $\pm$0.03-pixel manufacturing irregularity) and thus derive for
each broad-band UVIS filter a distortion solution that is accurate to
better than 0.01 pixel. As a by-product of the distortion solution,
\cite{2010AJ....140..631B} also gave an exciting foretaste of the
capabilities of WFC3 in identifying MPs in GCs.

In this paper we take the next quantum leap.  We distilled the
multitude of exposures taken for WFC3 calibration into a single,
comprehensive catalog. The large-dither and varied-orient observing
strategy adopted by the WFC3 team to pin down systematic issues is
also ideal for studying the cluster scientifically.  We undertook a
comprehensive reduction of the WFC3/UVIS and IR archive to produce a
26-band catalog with superlative photometry. When available, stars in
the photometric catalog are supplied with the high-precision
proper-motion (PM) measurements of \cite{2014ApJ...797..115B}.

This is the first of a series of several papers; in this paper we
describe in detail the data reduction processes and provide the
astro-photometric catalog to the astronomical community.  Future
papers in this series will be primarily focused on: (i) a
comprehensive analysis and characterization of the MPs of the cluster
in all different evolutionary sequences, from both the photometric and
kinematic point of views; (ii) continuing the search for a central
massive object; and (iii) searching for cataclysmic variables and He
white dwarfs.

This paper is organized as follows. In Section~\ref{sec:2}, we present
the data sets used for our study. Sections~\ref{sec:3} and \ref{s:cal}
are focused on the detailed description of the photometric reduction
and its calibration. In section~\ref{s:qpar}, we introduce several
photometric-quality parameters that can be used to identify and remove
poorly-measured objects. Section~\ref{s:comp} is dedicated to
artificial-star tests, while Section~\ref{s:astro} summarizes the
astrometric registration of stellar positions on the Gaia reference
system. Proper motions are discussed in Section~\ref{s:pm}. We
describe the assembly of the final astro-photometric catalog in
Section~\ref{s:cat}. Finally, we end with a summary of upcoming
scientific results and additional studies (Section~\ref{s:summ}).

\section{Data sets}
\label{sec:2}

The core of \wcen\ has been observed through many of the WFC3 filters
since 2009 for calibration purposes, and new observations continue to
be scheduled.  Table~1 summarizes the massive archive data, organized
in a camera/filter fashion. We downloaded from the archive a total of
655 exposures ($\sim$205 Ks)\footnote[3]{Those available at the time
  we undertook this project.}, taken through 26 different bands:\ 18
for WFC3/UVIS (385 exposures) and 8 for WFC3/IR (270 exposures).

For most of these filters, the central cluster region has been
observed through 18 to 62 completely independent, widely-spaced
pointings, and at 5 to 10 different roll angles. While these images
easily met the Institute's formal goals for distortion, flat-field
calibration, and stability monitoring, they are also well-suited to
higher-precision analysis.

Figure~\ref{f:fovs} shows the footprints of the WFC3 data set relative
to the cluster's center, divided by filter.  We color-coded the
footprints, from purple to brown, as a function of the filters'
central wavelength, from 225 to 1600 nm.  The dotted circle in each
panel highlights the cluster's core radius $r_{\rm c}=2\farcm37$
(\citealt{1996AJ....112.1487H}, 2010 edition). The transmission curve
of each filter, as a function of wavelength, is shown in the top panel
of Fig.~\ref{f:tc}.  The three bottom panels of the figure are a
zoomed-in view around UVIS medium- and narrow-band filters.

To give a visual sense of the breadth offered by this unique data set,
we show in the larger panel of Fig.~\ref{f:tiles} the trichromatic
image stack (red=F814W, green=F606W, blue=F275W) of a
$20^{\prime\prime}$$\times$$20^{\prime\prime}$ region near the
cluster's center. Several bright blue (HB) and red (RGB) stars
dominate the scene, surrounded by fainter MS stars of white hue. The
other panels of the Figure show the same region as seen through the
monochromatic image stacks of a selection of 14 filters, from the UV
(F225W) to the IR (F160W). In the first and the last of these smaller
panels we highlight HB and RGB stars with blue and red circles,
respectively. Moving from the bluer to the redder filters, HB stars
dim while RGB stars become the most prominent sources.

A wider picture of the center of the cluster
(3$^\prime$$\times$$2\farcm5$) is given in Figure~\ref{f:trichro}. In
order to maximize the color extension of the image, we chose here
F225W for the blue channel, F160W for the red channel, and the
intermediate F814W for the green channel.\footnote[4]{The green
  ``dot'' that can be spotted in the center of the brightest red stars
  is due to the behavior of the WFC3/IR detector when dealing with
  saturated pixels.} Orientation and scale are also shown on the
bottom-right corner. Different filter combinations allow the user to
highlight different cluster features, and some color/channel
combinations work better than others for different purposes. The
figure here is just an example. We are publishing a stack image for
all 26 filters, so that the interested reader has access to the full
range of wavelengths.

\subsection{WFC3/UVIS}
\label{sec:2.1}

The WFC3/UVIS data come from 12 different calibration programs,
specifically:\ flat-field uniformity (CAL-11452, PI:\ Quijano),
filter-dependent, low-frequency flat-field (L-flat) corrections
(CAL-11911 \& CAL-12339, both PI:\ Sabbi), fringe calibration
(CAL-11922, PI:Sabbi, \& CAL-12091, PI:\ Wong), geometric distortion
(GD) corrections (CAL-11911, PI:\ Sabbi, \& CAL-12353,
PI:\ Kozhurina-Platais), GD stability (CAL-12714 \& CAL-13100, both
PI:\ Kozhurina-Platais), image skew (CAL-12094, PI:\ Petro),
saturation-induced IR persistence (CAL-12694, PI:\ Long), count-rate
non linearity range and precision (CAL-12700, PI:\ Riess), and
post-flash characterization (CAL-12802, PI:\ MacKenty).

\subsection{WFC3/IR}
\label{sec:2.2}

The WFC3/IR data also come from different calibration programs --7 in
this case--, three of which are in common with WFC3/UVIS. Unique
programs are:\ low-frequency flat and GD correction (CAL-11928,
PI:\ Kozhurina-Platais), L-flat correction (CAL-12340, PI:\ Dahlen),
and saturation-induced IR persistence (CAL-12351, PI:\ Long). The
other three programs are designed to correct the GD: CAL-12353,
CAL-12694, and CAL-12714.

\section{Data Reduction}
\label{sec:3}

We used only \texttt{\_flt}-type exposures for this project, as they
preserve the un-resampled pixel data for stellar-profile fitting.  All
WFC3/UVIS exposures were corrected to remove charge-transfer
efficiency (CTE) defects\footnote[5]{CTE defects are caused by the
  presence of charge traps in the detectors that capture electrons,
  which are then released at a delayed time during the read-out
  process.}, following the empirical pixel-based approach described in
detail in \cite{2010PASP..122.1035A}.  Since the pixels of the WFC3/IR
detector are connected to capacitors, and IR exposures are read out in
a non-destructive, multi-plexed mode, no CTE correction is necessary.
In the following, we will illustrate the two-step procedure that we
used to measure stellar positions and fluxes.

\subsection{First-Pass Photometry}
\label{ss:fpp} 

\begin{figure}[!t]
\centering
\includegraphics[width=4.25cm]{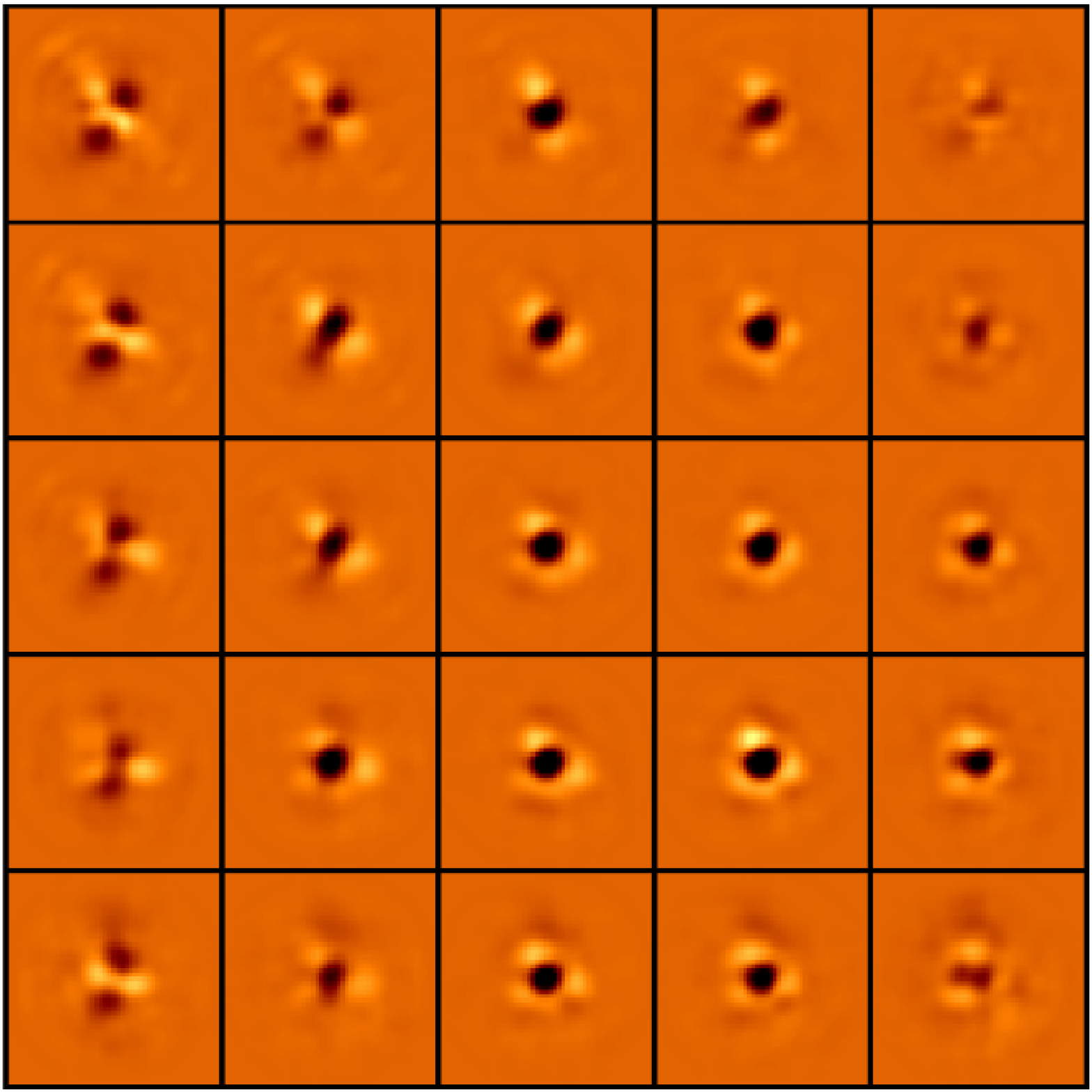}$\!$
\includegraphics[width=4.25cm]{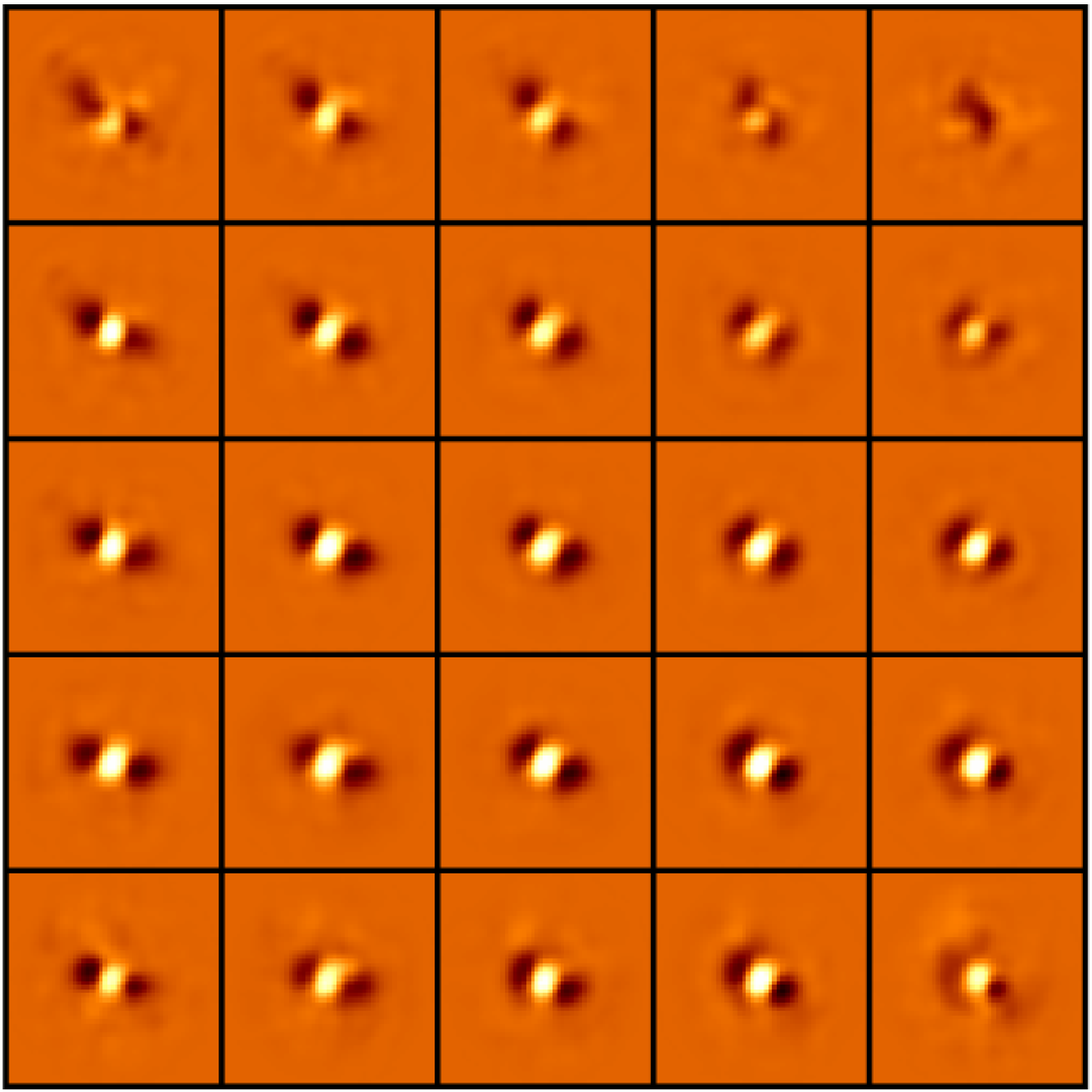}
\caption{\small{$5\times5$ Perturbation-PSF arrays obtained from two
    consecutive F275W exposures taken within the same orbit. The
    maximum peak-to-peak variation between these 2 sets of
    perturbation PSF models is $\sim6$\% with respect to the library
    PSF. (But it can be as high as 10\%).}}
\label{f:psf}
\end{figure}

The stable environment of space makes \textit{HST} an excellent
photometric and astrometric tool. In particular, \textit{HST}'s
point-spread functions (PSFs) have been extremely stable for the past
25 years. Therefore, a time-averaged empirical ``library'' PSF is
typically sufficient for most scientific projects. However, for
high-precision astrometry and photometry, it is important to take into
account even the smallest PSF variations, e.g. those induced by the
so-called telescope breathing. We have developed a new PSF-modeling
technique for both the UVIS and the IR channels that creates an array
of time-dependent, spatially-varying empirical PSFs for each
individual exposure.

To generate our improved PSF models we proceeded as follows.  We
started by selecting only bright, unsaturated, and isolated stars in
each exposure (there are plenty of stars with these properties in the
core of \wcen), and subtracted them using the appropriate,
state-of-the-art, spatially-varying (but time-constant) ``library''
PSF for each
filter\footnote[6]{{http://www.stsci.edu/$\sim$jayander/WFC3/}.}.
Subtraction residuals were then collected and averaged into an array
of perturbation PSFs (5$\times$5 for UVIS, 3$\times$3 for IR),
sampling each region of the detectors.  We explored the possibility of
using different array sizes, but we found that the 5$\times$5 UVIS and
the 3$\times$3 IR arrays are sufficient to fully account for
time-dependent variations across the image\footnote[7]{Note that the
  library UVIS PSFs are in a $7\times8$ array format, therefore the
  detector-related PSF spatial variations are already taken into
  account. The library IR PSFs consist of only one PSF for the entire
  detector, so our time-dependent perturbation PSFs will also
  automatically take into account for possible spatial variations.}.

To give a sense of the size of time-dependent PSF variations, we show
in Fig.~\ref{f:psf} two of these $5\times5$ perturbation arrays,
obtained from two consecutive F275W exposures. We used the same linear
scale for both sets. A lighter (darker) color means more (less) flux
with the respect to the library PSF.  The figure shows substantial PSF
variations across the detector with respect to the reference library
PSF, as well as from one exposure to the the next. Typically, between
5\% and 10\% of the total PSF flux can be redistributed within its
pixels from one exposure to the next.

With the new sets of PSFs so derived, we measured positions and fluxes
of stars in each individual exposure using the publicly-available
\texttt{FORTRAN} programs \texttt{img2xym\_wfc3uv} and
\texttt{img2xym\_wfc3ir}, directly adapted from the ACS/WFC
\texttt{img2xym\_WFC} package (\citealt{ak06}). These programs find
and measure stars in a single pass in each individual exposure,
without neighbor subtraction.

\subsection{The Master Frame}
\label{ss:master} 

Star positions were then corrected using the state-of-the-art GD
solutions of \cite{2011PASP..123..622B} for UVIS, and those made by
J.\ Anderson for the IR
channel\footnote[8]{\url{http://www.stsci.edu/$\sim$jayander/STDGDCs/}.}).
These GD corrections allow us to bring each exposure into a
distortion-free frame to better than the $\sim0.01$-pixel level.

We created a master-frame list based on the WFC3/UVIS F606W
single-exposure catalogs, since they are the most numerous and cover
the largest FoV. We adopted the X,Y positions of stars measured in the
first F606W exposure (\texttt{ibc301qrq\_flc.fits}) as a reference,
and used general, six-parameter linear transformations to
cross-identify stars in the reference frame with those measured in the
other F606W exposures.  Stellar positions and fluxes on the master
list are obtained as the 2.5$\sigma$-clipped average of the individual
measurements in each exposure. The master frame extends over about
8500$\times$9000 UVIS pixels ($\sim$340$\times$360 square arcsecs),
and has a pixel scale of $0\farcs04$. The cluster's center on the
master frame is approximately located at pixel position $(4300,4990)$.

Single-exposure catalogs of the remaining filters were all transformed
and averaged into the reference system defined by the master frame by
means of six-parameter linear transformations.  The single-filter,
average catalogs obtained this way constitute our ``first-pass''
photometry. The photometric zero-points of these catalogs are tailored
to those of the first (oldest) deep exposures of each filter (which
are also typically the most central ones).

\subsection{Second-pass photometry}
\label{ss:spp} 

The second-pass photometry allows us to find and measure stars in all
the individual exposures for the entire set of filters,
simultaneously. To this aim, we made use of the \texttt{FORTRAN}
software package \texttt{kitchen\_sync2} (KS2, Anderson in
preparation, see also \citealt{2016ApJS..222...11S}, their Section~3,
for an in-depth description of the software).  This package is a
generalization of the software developed to reduce the ACS Globular
Cluster Treasury project of GO-10775 (Pi:\ Sarajedini), described in
detail by \cite{2008AJ....135.2055A}.  The new version of the software
is able to handle multiple filters, large numbers of exposures, and
also is able to find and measure faint stars that cannot be detected
in individual exposures.

In brief, the routine takes the results of the first-pass photometry
cross-matched with the master list in order to define the
transformations (astrometric and photometric) from each exposure into
the reference frame.  It then goes through the field one patch at a
time (where a patch is a square region 125 pixels on a side) and uses
all the exposures together to find and measure the stars.  Since we
have a large number of filters, many of which have low throughput, we
decided to do our ``finding'' using only the F606W and F438W
exposures, which allow us to find reasonably faint stars along both
the MS and the WD cooling sequence.

KS2 finds stars in multiple passes over each patch.  It starts with
knowledge of the extremely bright (often saturated) stars in the field
along with a rough model of the extended PSF.  This allows it to know
which features might be PSF halos or diffraction spikes, so that it
will minimize the inclusion of spurious objects in the final list.
During its initial finding pass, the routine first identifies the
bright stars, then subtracts them to search for fainter stars in the
residuals in subsequent passes.  It has parameters that tell it how
well stars can be subtracted in order to prevent it from identifying
spurious sources in imperfectly-subtracted bright stars.

For this project, we had KS2 execute eight passes of finding.  The
last two passes were designed to find stars that are too faint to be
distinctly detected in individual exposures.  Since WFC3/UVIS images
are undersampled, the most we can hope for from the faintest stars is
that their contribution will push their centermost pixel above the
background in some fraction of images.  To find them, we made a map of
the patch, summing up in each pixel the number of exposures that had a
non-artifact-related local maximum in that pixel.  We compared the
number of marginal detections in each pixel against the number of
exposures available, and identified a star in pixels where there were
a significant number of detections.  For example, we would expect
random noise to produce a peak in 1 pixel out of 9.  If we have 90
exposures, we would expect there to be a noise peak in 10 of them.  If
we detect peaks in 25 of the exposures, then this represents a
significant event and we can confidently identify a source at that
location. In each wave of finding, KS2 identifies stars that satisfy
various criteria:\ isolation within a certain number of pixels;
significance level over the sky noise; quality of PSF fit; and number
of coincident peaks in multiple exposures. These criteria are set to
be increasingly more relaxed from the the first to the last iteration.

Most stars have rather different signal-to-noise levels in F606W and
F438W exposures. Were we to require a star to be found in both filters
simultaneously, we would have introduced significant selection
effects. Therefore, the first seven passes of finding were performed
on F606W exposures only, in order to maximize the chances of detecting
stars on the faint MS. The last pass was done only on F438W exposures,
in order to also detect reasonably faint WDs, which are typically 3
magnitudes fainter in F606W than in F438W at a given F438W$-$F606W
color with respect to faint MS stars.

KS2 has three approaches for measuring stars.  The first approach can
only be applied when a star generates a distinct peak within its local
5$\times$5-pixel, neighbor-subtracted raster.  When that happens, we
can measure a position and a flux for the star using the PSF
appropriate for that location in that exposure.  A sky value is
generally measured using surrounding pixels between 5 and 8 pixels
radius, with the contributions of the neighbors and the star itself
subtracted.  The brightest, unsaturated stars can be measured well
with the first approach, but fainter stars often do not produce a
significant peak in every exposure.  If we measure stars only when
they produce a peak, we will get a biased result.  For them, we have
devised methods two and three.  Method two takes the position
determined in the finding stage and uses that position and the PSF to
determine a best-fit flux from the inner 3$\times$3 pixels.  Method
three is similar, but it uses the brightest 4 pixels and weights them
by the expected values of the PSF in those pixels, which is
appropriate when the source is much fainter than the background
noise. The three approaches of measuring stars are best suited for
studies focused on different magnitude regimes. Stellar fluxes for the
three methods were obtained for the other filters by measuring
whatever flux was present in their exposures at the fixed locations
determined by the finding process.

For each filter, KS2 takes the positions and the fluxes (adjusted to
match the photometric zero point of the master frame for that filter)
measured for each star in each exposure and determines a robust
average for each star, along with an RMS of the residuals around the
mean that describes how consistently a star was measured in the
independent exposures.  Typically, all exposures taken through a given
filter have either the same exposure time, or the variation is less
than 30\%. Exposure-time variations of this size do not represent a
problem for KS2, because the signal-to-noise of a given star in those
exposures is still comparable.  There are three relevant
exceptions:\ F225W, F275W and F336W.  For these filters, short
exposures are considerably shorter than the long ones (from about half
to one tenth).  For these 3 filters, we let KS2 measure the photometry
only on the long exposures. Photometry of stars that are saturated in
the long exposures is recovered using the 1-pass photometry.

\begin{figure}[!t]
\centering
\includegraphics[width=\columnwidth]{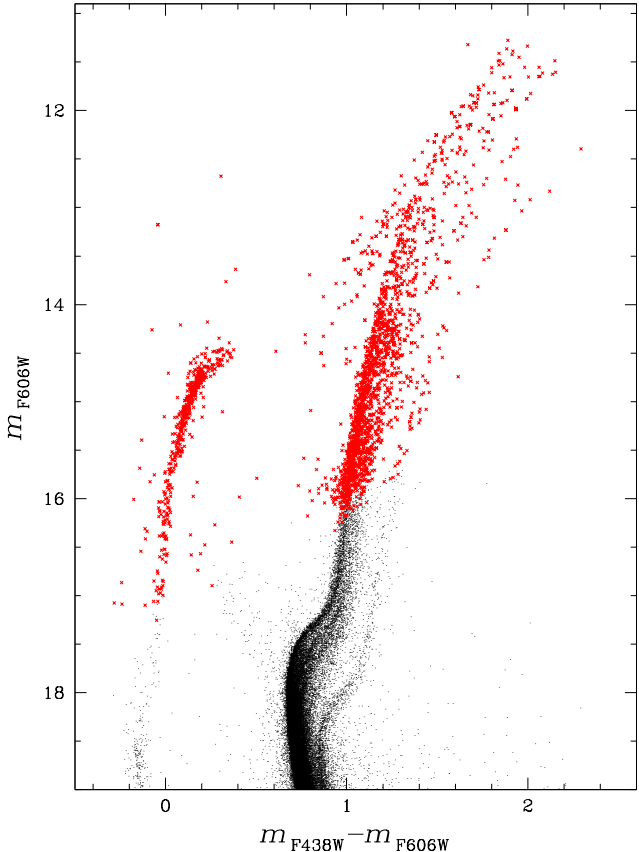}
\caption{\small{The bright part of the $m_{\rm F606W}$ vs. $m_{\rm
      F438W}-m_{\rm F606W}$ CMD. Unsaturated stars (in black) are
    directly measured by KS2, while the photometry of saturated stars
    (red crosses) come from the first-pass photometry.}}
\label{f:sat}
\end{figure}

Our final photometric catalog contains a total of 478\,477 sources and
includes results from all three methods for all 26 filters.

\subsection{Saturated stars}
\label{ss:ss}

Positions and fluxes of saturated stars are not fitted by
KS2. Instead, these measurements are available through the first-pass
photometry. KS2 makes use of positions and fluxes of saturated stars
to construct weighted masks around them, which help the software to
avoid PSF-related artifacts. As a result, we supplemented the output
of KS2 with saturated stars as measured by the first-pass photometry.
\cite{gill04} showed that the ACS detector conserves electrons even
when full-well saturation causes them to bleed from the place where
they were generated.  The same is true for WFC3/UVIS
(\citealt{gill10}).  Our \textbf{first}-pass software routine
(\texttt{img2xym}) is able to identify saturated stars and locate
their centers to an accuracy of about one pixel. It adds up all the
relevant flux from the bled-into pixels to determine an accurate flux
for each saturated star (mode details can be found in Sect.~8.1 of
\citealt{2008AJ....135.2055A}).  Figure~\ref{f:sat} shows the upper
part of the $m_{\rm F606W}$ vs. $m_{\rm F438W}-m_{\rm F606W}$ CMD of
\wcen. Black points are unsaturated stars as measured by KS2 using
method one. Stars saturated in either filter are marked with red
crosses.

The UVIS and IR channels, due to the different nature of their
detectors, deal with saturated pixels in a different way. The excess
flux in a UVIS pixel is simply ``bled'' onto adjacent, unsaturated
pixels along the Y axis, creating the so-called bleeding columns on
the image. On the other hand, IR detectors perform multiple readings
of the flux of a pixel at a different integration time. These values,
as a function of the integration time, are fit by a straight line, the
slope of which provides the total count rate of a pixel. When a pixel
reaches saturation before the end of the exposure, then only
unsaturated reads are used to estimate its flux.  Very bright stars on
IR images are easily identifiable because their central pixels have
fewer counts than the outer ones; this is due to their central pixels
reaching saturation before the first read.  Our first-pass photometry
measures positions and fluxes of saturated stars only in UVIS
exposures. As a result, our final IR photometry will not include
information about saturated stars.

\begin{figure*}[!t]
\centering
\includegraphics[height=4.6cm]{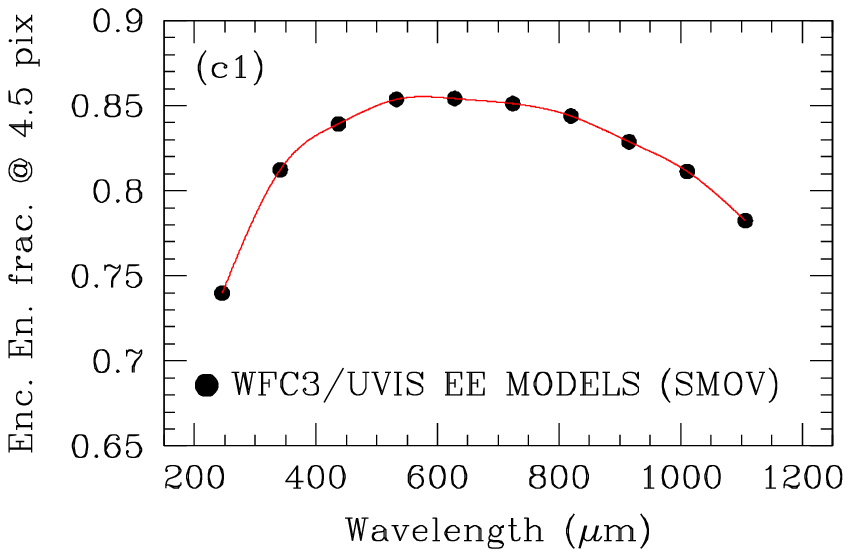}
\includegraphics[height=4.6cm]{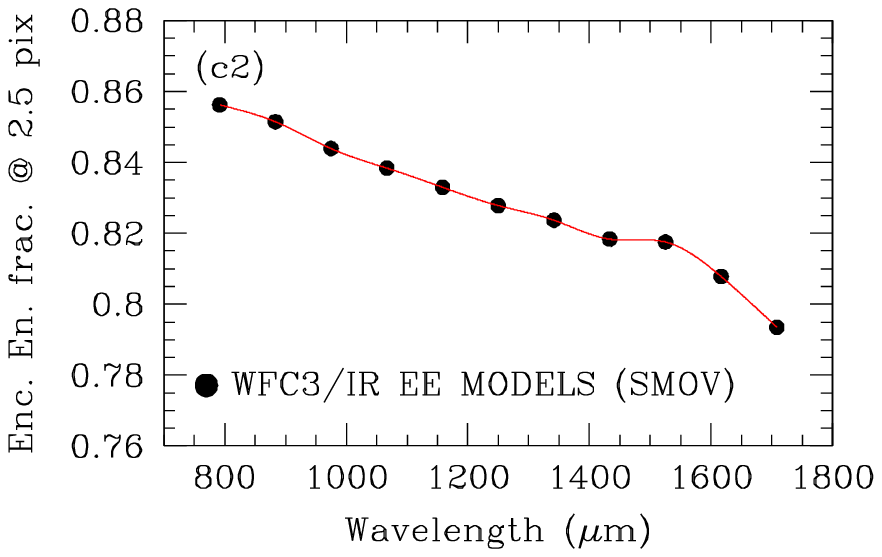}
\includegraphics[height=4.6cm]{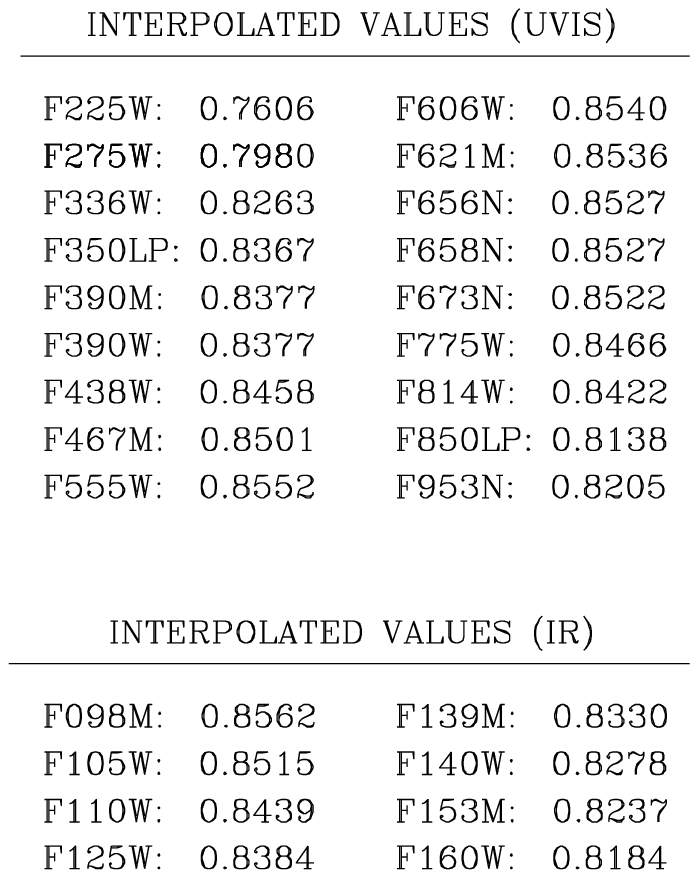}
\caption{\small{(\textit{Left:}) Interpolated WFC3/UVIS
    encircled-energy fractions as a function of the wavelength for a
    4.5-pixel aperture radius. (\textit{Middle:}) WFC3/IR
    Ensquared-energy fractions as a function of the wavelength for a
    2.5$\times$2.5-pixel$^2$ squared aperture. (\textit{Right:}) For
    completeness, we report the interpolated (and adopted) values for
    each of the 26 filters.}}
\label{f:tabval}
\end{figure*}

\subsection{Caveats}
\label{ss:cav}

Several considerations need to be kept in mind when using our
photometry. We have already discussed the complications with saturated
stars, but there are three remaining issues:\ (i) the lack of an
appropriate GD correction for some filters; (ii) the lack appropriate
library PSFs for some filters; and (iii) the effect of crowding (and
--to some extent-- persistence) in IR exposures.

In \cite{2011PASP..123..622B}, we derive state-of-the-art GD
corrections for 10 UVIS wide-band filters: F225W, F275W, F336W, F390W,
F438W, F555W, F606W, F775W, F814W, and F850LP.  In addition, we also
made use of the GD correction for the F467M filter, which was obtained
following \cite{2011PASP..123..622B} prescriptions, and using the
exquisitely-dithered exposures of the large \textit{HST} program
GO-12911 (PI:\ Bedin), see \cite{2013AN....334.1062B}.  At the time of
writing, there are no suitable exposures to properly model the GD for
the remaining UVIS filters employed here. For these filters, we simply
adopted the F606W GD correction, which worked fairly well, with
positional residuals no larger than a few \textbf{hundredths} of a
pixel. \textbf{Moreover, we found no strong dependence of the
  positional residuals as a function of the filter wavelength.}

A similar argument can be made for our library PSFs. We currently have
library-PSF models for the same 10 wide-band filters for which we have
a GD correction, plus F467M\footnote[9]{The library PSF for the F467M
  filter was also derived using GO-12911 exposures.}. We lack of
library PSFs for the following UVIS filters:\ F350LP, F390M, F621M,
F656N, F658N, F673N, and F953N. In addition, we don't have library PSF
models for the IR filters F105W, F140W, and F153M.  For those
exposures taken through these filters, we created our own
single-exposure PSF models by perturbing the available library PSFs of
the filter with the closest central wavelength. The photometry
obtained with these custom PSF models generally has larger errors and
worse \texttt{QFIT} values\footnote{The KS2 \texttt{QFIT} is defined
  as the linear correlation coefficient between the pixel values and
  the PSF model.}. The user should take extra care in analyzing
photometric data obtained for the filters that have no library PSFs.

The core of \wcen\ is only moderately crowded in UVIS exposures, but
the crowding level is dramatically higher in IR exposures, due to the
larger pixel size and worse undersampling. As a result, the
photometric quality of IR exposures is significantly lower than that
of UVIS exposures. We will include IR photometry information in our
catalog for the sake of completeness, however, we discourage the
reader from using the IR photometry for any but the brightest sources
(RGB) on account of the significant crowding in the coarse-resolution
IR images.  Probably the most useful products of our IR reduction are
the high-quality image stacks.

\section{Photometric calibration}
\label{s:cal} 

The process of calibrating --a more appropriate description would be
zero-pointing-- \textit{HST}'s photometry is based on the comparison
between our PSF-based instrumental magnitudes, measured on
\texttt{\_flt} exposures (CTE-corrected for UVIS), and the
aperture-photometry-based magnitudes as measured on the \texttt{\_drz}
exposures (which are always normalized to 1-second exposure time):
$$ 
{\rm CAL_{\rm{filter}}} = {\rm INSTR}_{\rm filter}^{\rm flt} +
\Delta{\rm mag} + {\rm ZP}_{\rm{filter}}
$$
where ${\rm CAL_{\rm{filter}}}$ is the calibrated photometry for a
particular filter; ${\rm INSTR}_{\rm filter}^{\rm flt}$ is our
instrumental photometry as measured on \texttt{\_flt} exposures;
$\Delta{\rm mag}$ is the median magnitude difference between ${\rm
  AP}_{r,\infty}^{\rm drz}(\lambda)$, the aperture photometry measured
on \texttt{\_drz} exposures within a finite-aperture radius $r$, but
corrected to account for an infinite-aperture radius, and our
instrumental magnitudes\footnote[10]{To this aim, we used the
  \texttt{\_drz} images relative to the same exposures used to define
  the photometric zero points of the single-exposure catalogs.}; and
finally ${\rm ZP}_{\rm{filter}}$ is the filter zero point in a given
photometric system.  The ${\rm ZP}_{\rm{filter}}$ values are tabulated
--and monitored-- on the STScI
webpage\footnote[11]{\url{http://www.stsci.edu/hst/wfc3/phot\_zp\_lbn}.}.
Please note that WFC3/UVIS exposures are in electrons, while WFC3/IR
exposures are in electrons per second. As such, when we compare our
UVIS ${\rm INSTR}_{\rm filter}^{\rm flt}$ magnitudes to ${\rm
  AP}_{r,\infty}^{\rm drz}(\lambda)$ ones, we are also naturally
accounting for the exposure-time term.

We used only a subset of bright, unsaturated, relatively isolated
stars (no brighter neighbors within 20 pixels) for the photometric
calibration of each filter. Our photometry is calibrated to the
VEGA-MAG system.

In order to obtain ${\rm AP}_{r,\infty}^{\rm drz}(\lambda)$, we need
to choose a suitable, \textit{finite} aperture radius ${r}$ with which
to measure stellar photometry on \texttt{\_drz} exposures, and then
apply the proper filter-dependent aperture
correction\footnote[12]{Encircled and ensquared energy fractions as a
  function of $r$, for tabulated wavelengths, can be found on the WFC3
  webpage: \url{http://www.stsci.edu/hst/wfc3/}.}.  The larger the
aperture radius we choose, better constrained the aperture correction
will be. On the other hand, in relatively crowded environments such as
the core of \wcen, a small aperture radius clearly helps us in
minimizing the light contamination from neighbor sources.

\begin{table}[!t]
\label{tab:zps}
\centering
\small{
\begin{tabular}{ccc}
\multicolumn{3}{c}{\textsc{Table~2}}\\
\multicolumn{3}{c}{\textsc{Photometric-calibration}}\\
\multicolumn{3}{c}{\textsc{zero points}}\\
\hline\hline
\textbf{Filter}&\textbf{$\Delta$mag}&\textbf{ZP (VEGA)}\\
               & (mag)                &(mag)\\
\hline
\multicolumn{3}{c}{WFC3/UVIS}\\
\hline
F225W & +7.0889 $(^{-0.2967}_{+7.3856})$  & +22.3808\\
F275W & +7.0240 $(^{-0.2337}_{+7.2577})$  & +22.6322\\
F336W & +6.1648 $(^{-0.1954}_{+6.3602})$  & +23.4836\\
F350LP& +6.1925 $(^{-0.1677}_{+6.3602})$  & +26.7874\\
F390M & +6.1867 $(^{-0.1735}_{+6.3602})$  & +23.5377\\
F390W & +6.2007 $(^{-0.1595}_{+6.3602})$  & +25.1413\\
F438W & +6.1961 $(^{-0.1641}_{+6.3602})$  & +24.9738\\
F467M & +6.4827 $(^{-0.1503}_{+6.6330})$  & +23.8362\\
F555W & +3.8302 $(^{-0.1749}_{+4.0051})$  & +25.8160\\
F606W & +3.8517 $(^{-0.1534}_{+4.0051})$  & +25.9866\\
F621M & +6.9822 $(^{-0.1429}_{+7.1251})$  & +24.4539\\
F656N & +6.6090 $(^{-0.1384}_{+6.7474})$  & +19.8215\\
F658N & +6.2200 $(^{-0.1402}_{+6.3602})$  & +20.6795\\
F673N & +6.2313 $(^{-0.1289}_{+6.3602})$  & +22.3297\\
F775W & +6.4621 $(^{-0.1709}_{+6.6330})$  & +24.4747\\
F814W & +3.8445 $(^{-0.1616}_{+4.0051})$  & +24.6803\\
F850LP& +4.2503 $(^{-0.1951}_{+4.4454})$  & +23.3130\\
F953N & +7.1632 $(^{-0.1603}_{+7.3235})$  & +19.7549\\
\hline
\multicolumn{3}{c}{WFC3/IR}\\
\hline
F098M & $-$0.0736 & +25.1057\\
F105W & $-$0.0335 & +25.6236\\
F110W & $-$0.0743 & +26.0628\\
F125W & $-$0.0844 & +25.3293\\
F139M & $-$0.0633 & +23.4006\\
F140W & $-$0.1262 & +25.3761\\
F153M & $-$0.0636 & +23.2098\\
F160W & $-$0.0756 & +24.6949\\
\hline\hline
\end{tabular}}
\end{table}

To select the best value for aperture radius in our field, we measured
the \texttt{\_drz} aperture photometry of bright, relatively isolated
and unsaturated stars using 10 different apertures:\ 2.5, 3.0, 3.5,
4.0, 4.5, 5.0, 5.5, 6.0, 7.5, and 10 pixels.  We used circular
apertures for WFC3/UVIS exposures, and squared apertures for WFC3/IR
exposures.  A local sky value was computed between 12 and 16 pixels in
all cases.  Each of these measurements was then properly corrected to
account for the finite aperture. To do so, we had to interpolate the
aperture-correction values, tabulated for a subsample of different
apertures and central wavelengths\footnote[13]{The tabulated values
  are also available at \url{http://www.stsci.edu/hst/wfc3/}.}.

We cross-identified stars in common between the \texttt{\_drz}-based
aperture photometry and our KS2 method-one photometry, and computed
the 2.5$\sigma$-clipped median values $\Delta {\rm mag}={\rm
  AP}_{r,\infty}^{\rm drz}(\lambda)-{\rm INSTR}_{\rm filter}^{\rm
  flt}$ for the 10 different aperture radii. The number of rejected
measurements started to dramatically increase for aperture radii
larger than 4.5(2.5) pixels for all UVIS(IR) filters, due to neighbor
contamination.  As a best trade-off between the need of large
apertures and solidly determined $\Delta {\rm mag}$ values, we adopted
the aperture values of 4.5 and 2.5 pixels for UVIS and IR,
respectively.

Figure~\ref{f:tabval} shows the interpolation curves we obtained for
WFC3/UVIS (left) and WFC3/IR (middle) at the fixed apertures of 4.5
pixels and 2.5 pixels, respectively, as a function of the
wavelength. For completeness, on the right we report the adopted
correction values for each of the 26 filters.

\begin{figure*}[!t]
\centering \includegraphics[width=\textwidth]{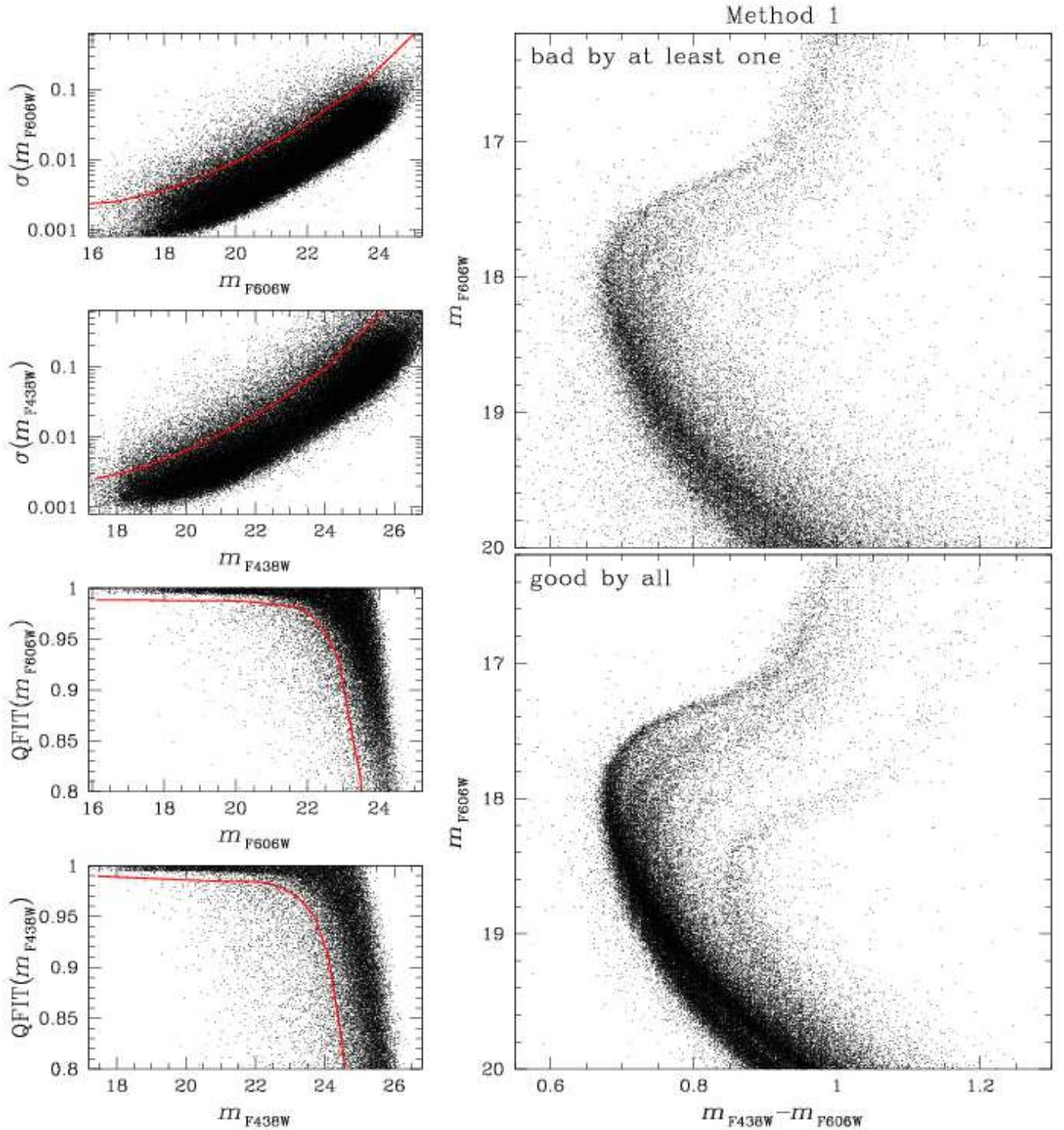}
\caption{\small{The effects of simple stellar selections based on
    photometric errors and \texttt{QFIT} parameters on the method-one
    $m_{\rm F606W}$ vs. $m_{\rm F438W}-m_{\rm F606W}$ CMD.  In the
    left panels we show, from top to bottom, $\sigma$($m_{\rm
      F606W}$), $\sigma$($m_{\rm F438W}$), \texttt{QFIT}($m_{\rm
      F606W}$) and \texttt{QFIT}($m_{\rm F438W}$) as a function of
    magnitude. The red lines (drawn by hand) separate the bulk of
    well-measured stars from those that are less well measured. The
    top CMD on the right show stars that failed at least one of the
    selection cuts. The bottom CMD contains only stars that appear to
    be well measured according to all the four parameters.}}
\label{f:m1sel}
\end{figure*}

\begin{figure*}[!t]
\centering
\includegraphics[width=\textwidth]{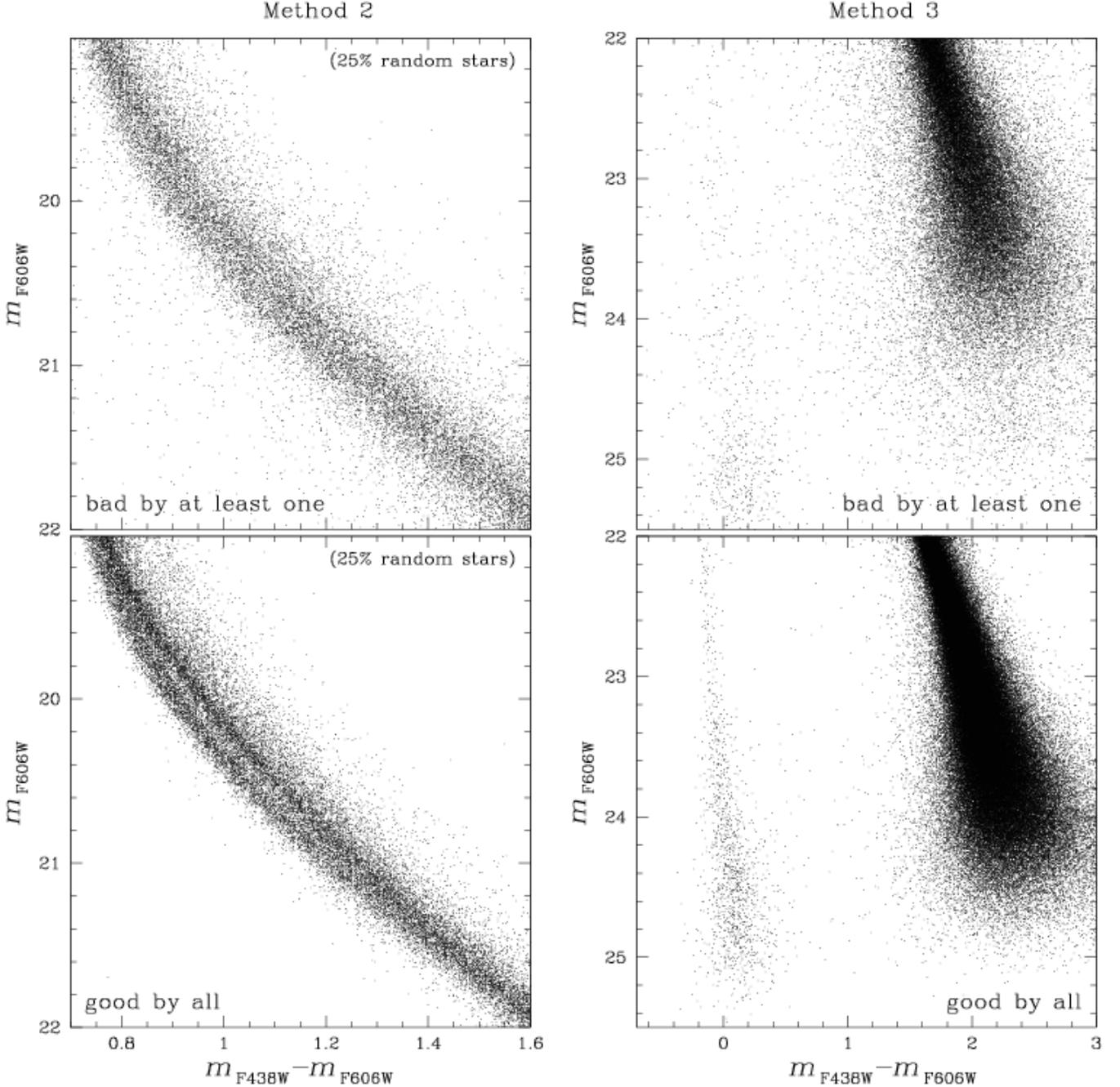}
\caption{\small{Very similar selection cuts to those shown in
    Fig.~\ref{f:m1sel} are applied to method-two (on the left) and
    method-three (on the right) photometries.  The top panels show
    those stars rejected by at least one of the four selection
    criteria, while the bottom panels contain only stars that passed
    all four cuts.  For clarity, we only plotted 25\% randomly
    selected stars on the left panels.}}
\label{f:m23sel}
\end{figure*}

\begin{figure*}[!t]
\centering
\includegraphics[width=\textwidth]{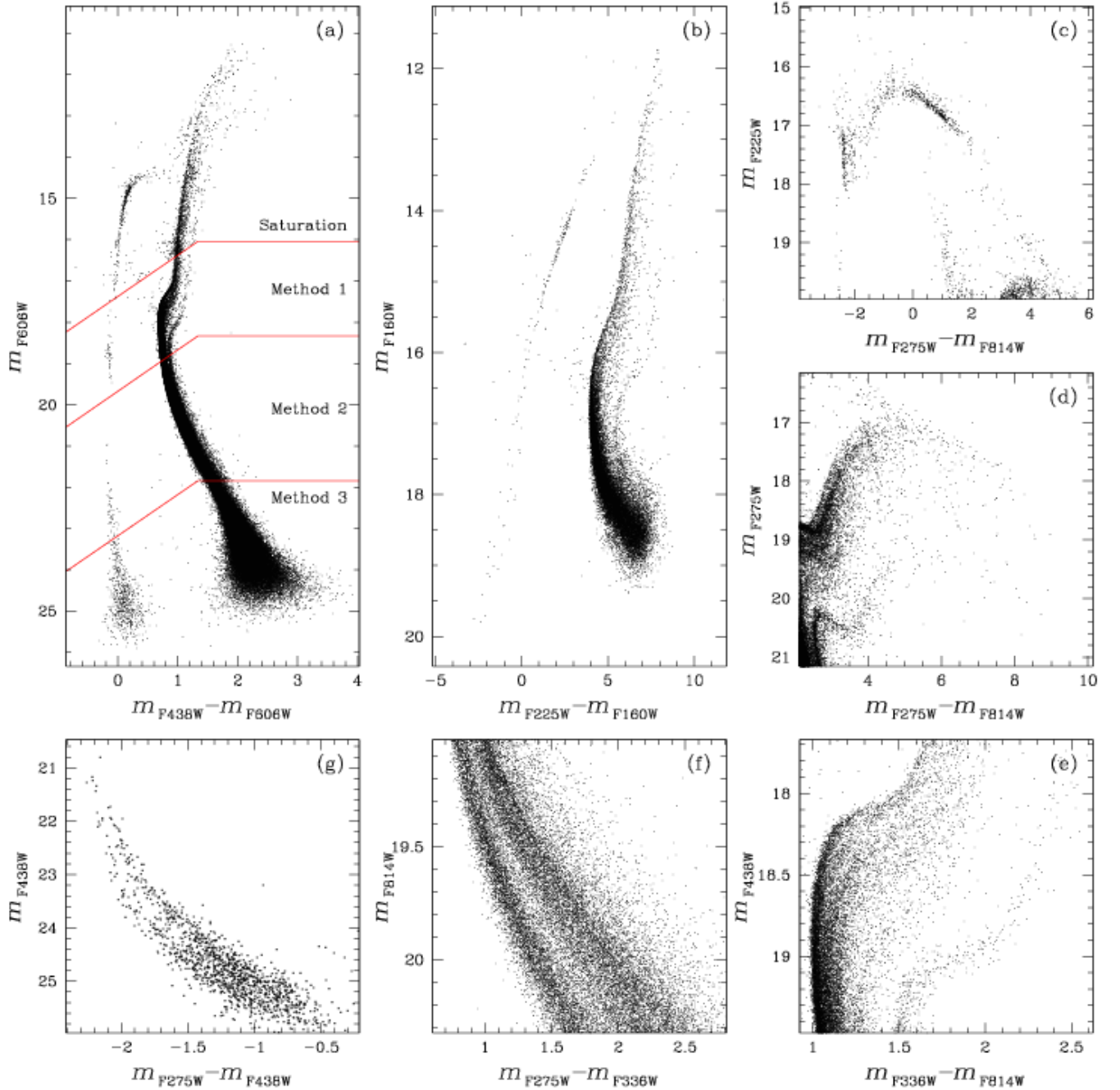}
\caption{\small{(a) The full $m_{\rm F606W}$ vs. $m_{\rm F438W}-m_{\rm
      F606W}$ CMD of \wcen\ obtained by combining together the best
    85\% of the stars measured in the three methods at any given
    magnitude. The transition between each photometric method is
    highlighted by red lines. Saturated stars are shown without any
    cut.  The same selection criteria are applied to the other panels
    of the Figure. (b) The $m_{\rm F225W}$ vs. $m_{\rm F225W}-m_{\rm
      F160W}$ CMD, which is the one with the widest possible color
    baseline obtainable with our photometry. Panels (c) to (g) provide
    an overview of the different evolutionary sequences of the cluster
    as seen through CMDs based on different filter
    combinations. Moving clockwise:\ (c) the HB, (d) the RGB, (e) the
    SGB, (f) the MS, and (g) the WD cooling sequence.}}
\label{f:ov}
\end{figure*}

\begin{figure*}[!t]
\centering
\includegraphics[width=\textwidth]{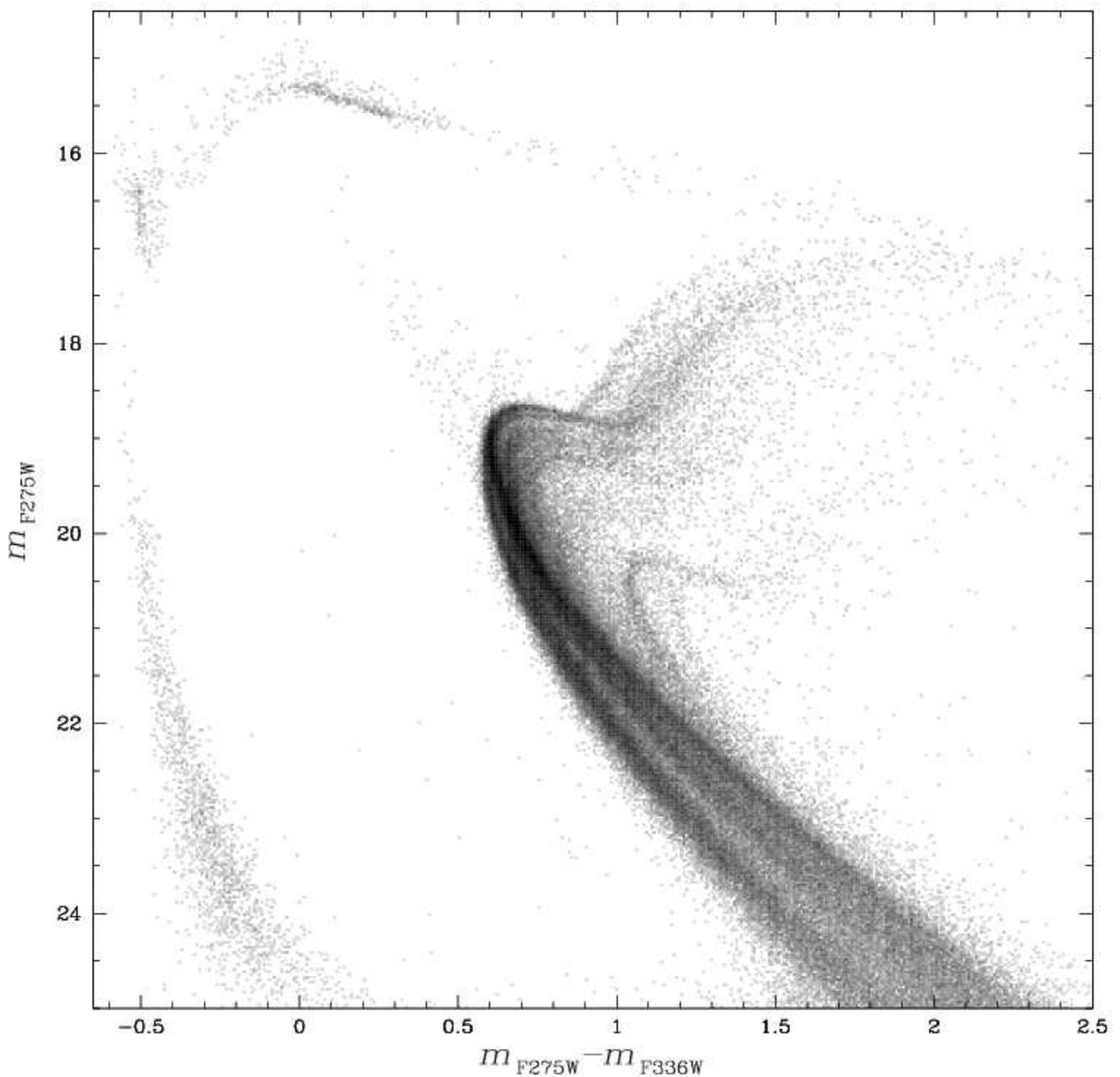}
\caption{\small{The Hess diagram of the $m_{\rm F275W}$ vs. $m_{\rm
      F275W}-m_{\rm F336W}$ CMD, showing the complex, majestic
    structure of the multiple stellar populations of \wcen.}}
\label{f:hess}
\end{figure*}

\begin{figure*}[!t]
\centering
\includegraphics[width=\textwidth]{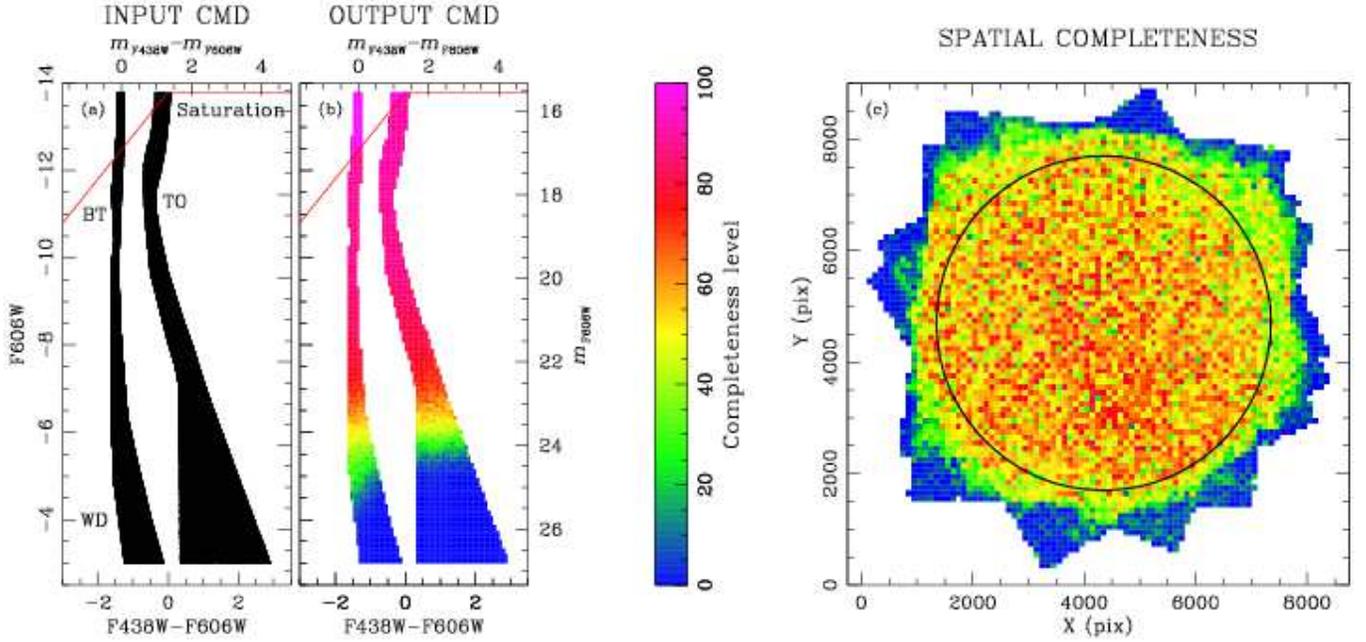}
\caption{\small{(a) Input CMD in both instrumental and calibrated
    magnitudes. The saturation level and a few evolutionary features
    (BT=blue tail, TO=turnoff, WD=white dwarfs) are also
    highlighted. (b) Hess diagram of the recovered method-two CMD,
    color-coded according to the completeness level shown in the
    middle of the figure. (c) Hess diagram of the spatial completeness
    for method-two instrumental magnitudes in the range $-$4.5 to
    $-$6.5. The regions with worse completeness levels are those
    mapped by only a few single exposures (typically only one in the
    corners of the FoV), or in close proximity of saturated stars.}}
\label{f:as1}
\end{figure*}

\begin{figure}[!t]
\centering
\includegraphics[width=\columnwidth]{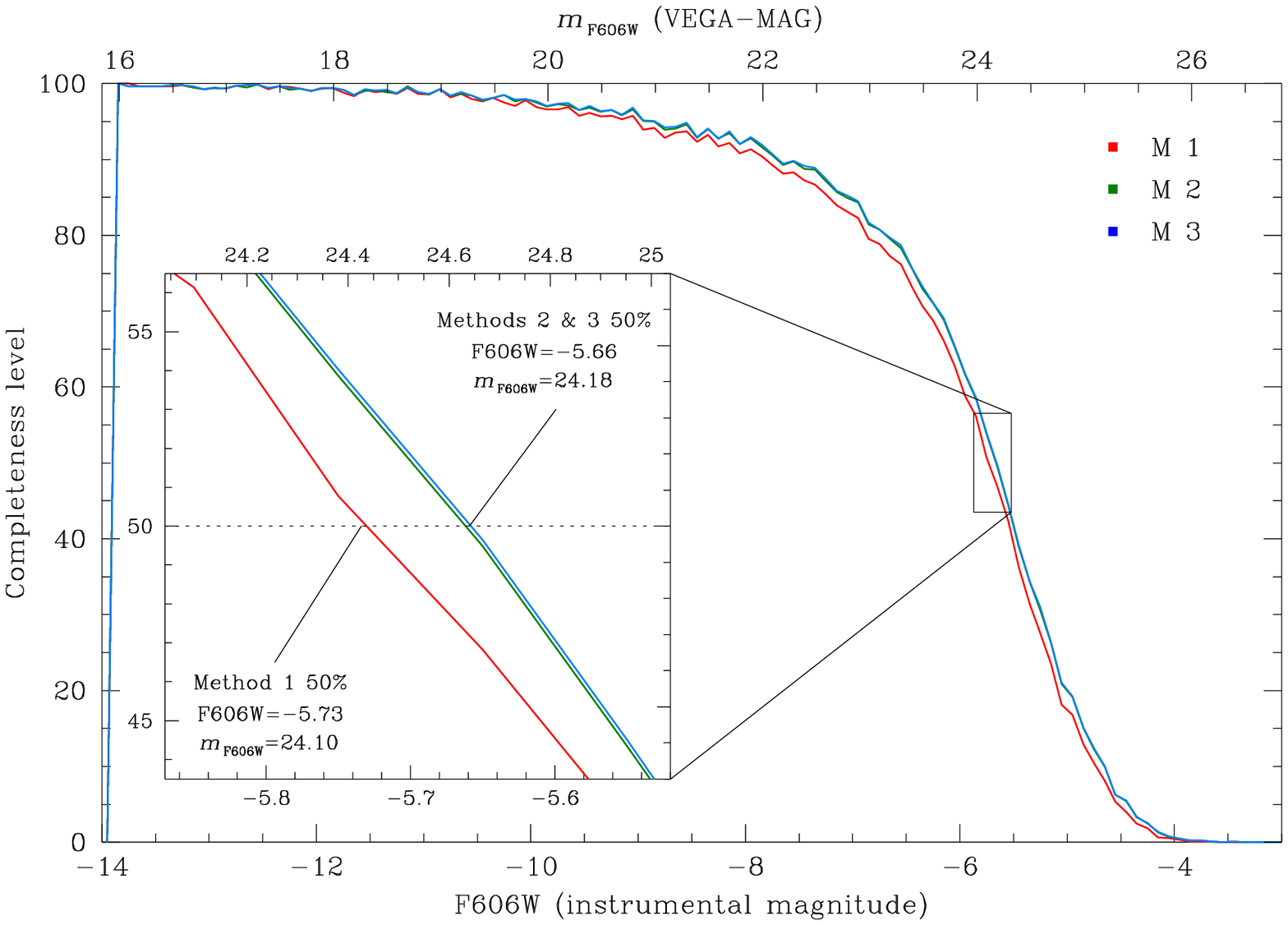}
\caption{\small{Completeness level as a function of the F606W
    magnitude (both instrumental and calibrated) for the three
    photometric methods (red for method one, green for method two, and
    blue for method three). The inset shows a zoom-in around the 50\%
    completeness level.}}
\label{f:as2}
\end{figure}

The photometric zero-point differences between each of the three KS2
methods, measured in the appropriate magnitude intervals where each
method is designed to work best, are always found to be less than
0.005 mag. Therefore, we simply applied the calibration corrections
obtained for method one to the other two methods.

Table~2 collects the aperture-correction $\Delta {\rm mag}$ values
obtained for each filter, together with the respective VEGA-MAG zero
points from the STScI website. For clarity, the WFC3/UVIS $\Delta {\rm
  mag}$ values are also split into two components (in
parentheses):\ the lower value is the exposure-time term ($+2.5\log
(t_{\rm exp})$), while the upper value is the offset between the
aperture-corrected \texttt{\_drz} magnitudes and our instrumental
magnitudes, corrected to a 1-second exposure.  There is no need to
split WFC3/IR $\Delta {\rm mag}$ into two components, since WFC3/IR
\texttt{\_flt} exposures are in units of electrons per second, and
therefore the exposure-time term is always zero.

These values allow the users to transform calibrated magnitudes back
into instrumental magnitudes\footnote[14]{Instrumental magnitudes are
  defined as $-2.5\log ({\rm total counts})$, where with ``total
  counts'' we mean the counts (in electrons or electrons per second,
  for UVIS and IR, respectively) under the fitted PSF.}, use different
aperture radii, and/or different photometric systems if their science
so requires.

By comparing our photometry with that in \cite{2010AJ....140..631B},
it is clear that the official zero points have changed since
(typically by a few hundredths of a magnitude). The photometry
presented here has the most-up-to-date calibration.

\section{Quality Parameters}
\label{s:qpar}

Different scientific goals are sensitive to different kinds of
photometric errors, and it is not trivial to come up with a single
parameter characterizing the photometric error for each
star. Furthermore, KS2 provides different ways of measuring the flux
of a source that are optimized for different signal-to-noise regimes.
To help us in selecting the best stars and best measurements for a
particular science case, KS2 provides additional quality parameters
other than just the agreement among independent measurements (RMS).

The quality of fit (\texttt{QFIT}) measurement can help to
discriminate between sources that are isolated and/or well measured by
the PSF, and sources for which the light profiles were not accurately
fit by the PSF (e.g., extended sources, blends, etc.).  Note that, in
contrast to all the other software-reduction packages produced by our
group, the KS2 convention for the \texttt{QFIT} values is:\ stars with
\texttt{QFIT} values close to one are well-fitted by the PSF, while
\texttt{QFIT} values close to zero imply poor measurements.

The parameter ``\textit{o}'' tells us the amount of neighbor flux that
was subtracted from the aperture before the target star could be
measured, divided by the flux of the target star itself.  Neighbor
subtraction can never be done perfectly, and it is hard to measure the
flux of faint sources surrounded by much brighter ones.  As such,
sources with large values of the \textit{o} parameter should be
considered suspect.

KS2 also reports the number of individual exposures in which a star
was found, $N_{\rm f}$, and the final number of good measurements
actually used to measure the star's flux, $N_{\rm g}$. Sources for
which a large fraction of measurements is discarded ($N_{\rm
  g}$/$N_{\rm f}$$\ll$$1$) should also be considered with suspicion.

Each of the three methods of measuring photometry works best in
different magnitude regimes, and comes with its own set of diagnostic
parameters. For example, method-1 is preferred for stars with
instrumental magnitudes\footnote[15]{instrumental magnitudes can be
  easily obtained from calibrated magnitudes by subtracting the values
  in columns 2 and 3 of Table~2.} brighter than --say--
$\sim$$-$11. Method two works best in the instrumental-magnitude range
between about $\sim$$-$11 and $\sim$$-$8, while method three should be
preferred for stars fainter than instrumental magnitude
$\sim$$-$8. The exact regimes within which one method should be
preferred over another varies for different filters and, all things
being equal, can vary for different scientific applications. As a
general rule of thumb, the user should favor the photometric method
that provides narrower sequences on the CMD.

As a simple, demonstrative example, in Figures~\ref{f:m1sel} and
\ref{f:m23sel} we show the effects of mild quality-parameter cuts on
the $m_{\rm F606W}$ vs. $m_{\rm F438W}-m_{\rm F606W}$ CMD.  In analogy
with what was done in \cite{2008AJ....135.2055A}, we based our
selections on four diagnostics (two per filter):\ the photometric
errors $\sigma$, defined as the photometric RMS divided by the square
root of $N_{\rm g}$, and the \texttt{QFIT} parameters.

The left panels of Fig.~\ref{f:m1sel} show, from top to bottom,
$\sigma$($m_{\rm F606W}$), $\sigma$($m_{\rm F438W}$),
\texttt{QFIT}($m_{\rm F606W}$) and \texttt{QFIT}($m_{\rm F438W}$) as a
function of their respective method-one magnitude. In each of these
panels we drew by hand a line (in red) separating the bulk of well
measured stars from a few outliers. The CMD on the top-right panel
shows those stars that do not qualify as well measured in at least one
of the cuts. The CMD on the bottom-right panel contains stars that
appear to be well measured according to all four parameters.  Most of
the poorly-measured sources populating the areas to the left and to
the right of the main sequence have been removed with these simple
cuts.

Figure~\ref{f:m1sel} focuses on method-one photometry, and the CMDs
are centered around bright, unsaturated stars. Figure~\ref{f:m23sel}
shows the effects of very similar selection cuts on the CMDs obtained
with the method-two (left) and method-three (right) photometric
algorithms. Again, stars that do not pass at least one of the
selection criteria are shown in the top panels, while only stars that
pass all four cuts are plotted in the bottom CMDs. Since method two
works best for stars of intermediate luminosity, the left CMDs are
zoomed-in around the central portion of the MS of \wcen. The bottom
part of the MS and the WD cooling sequence are instead shown using the
method-three photometry.  For clarity, only a randomly-selected 25\%
of the stars are shown in the left panels.  The bottom panels in the
Figure are clearly able to reveal the well-known double MS of the
cluster (left), as well as a well-defined and populated WD cooling
sequence (right).  Note that, for this demonstrative example, we
applied mild, arbitrary cuts. It stands to reason that more rigorous
selections should be applied, tailored to the user's particular
scientific needs, for high-precision photometric analyses.

Some studies might need to analyze stars spanning wide ranges in
magnitude, where two or more photometric methods work best at the same
time.  In these cases, different selection criteria might be more
desirable. For instance, one could choose to keep the same
best-measured fraction of the stars at any given magnitude.  As an
example, panel (a) of Fig.~\ref{f:ov} shows the full $m_{\rm F606W}$
vs. $m_{\rm F438W}-m_{\rm F606W}$ CMD of \wcen\ for stars that have
been detected in at least 10 individual exposures in both
filters. Only the best 85\% with respect to photometric errors and
\texttt{QFIT} parameters at any given magnitude level are shown.  This
way, we can combine the three different photometric methods together,
with a seamless transition between them.  No selection cuts were
applied to saturated stars. The three red lines in panel (a) define
the regions within which stars are either saturated in at least one
filter, or for which photometry is obtained with one of the three
methods. This CMD spans 15 magnitudes in $m_{\rm F606W}$, i.e. a
$10^6$ difference in flux between the faintest and the brightest
stars.

The remaining panels of Fig.~\ref{f:ov} offer a general overview of
our photometric catalog. In each of these panels, stars are selected
as in panel (a) for the appropriate filters.  In panel (b) we show the
$m_{\rm F225W}$ vs. $m_{\rm F225W}-m_{\rm F160W}$ CMD, which is the
one with the widest possible color baseline obtainable with our
photometry. Note a color difference of about 11 magnitudes between the
bluest and the reddest stars. The other panels in the figure are
focused on specific evolutionary sequences, and employ different
filter combinations. From panel (c) to (g), clockwise:\ the HB, the
RGB, the SGB, the MS, and the WD cooling sequences.

The careful reader might have noticed that the number of stars in the
double WD sequence of panel (g) is sizably lower than that on the CMDs
published in \cite{2013ApJ...769L..32B}. In
\cite{2013ApJ...769L..32B}, we tested a preliminary version of KS2 on
the three bluest filters F225W, F275W, F336W, plus F438W, with the aim
of analyzing in detail the structure of the WD cooling sequence. The
finding process was performed on F275W exposures only. This guarantees
us to find the faintest WDs in the data, at the cost of loosing many
faint redder sources on the MS. In this work, instead, our aim is to
produce a photometric catalog with as many reasonably faint stars on
\textit{both} sequences (WD and MS) as possible, and the finding
routine was run on F606W and F438W filters instead.

The Hess diagram shown in Fig. 10 of \cite{2010AJ....140..631B}
represents a sort of iconic picture of the complexity of the stellar
populations in \wcen. The photometric and astrometric quality of our
new data set is much improved with respect to that published in 2010,
and spurred us to produce the even more mesmerizing
Fig.~\ref{f:hess}. All the multiple-population features discussed in
\cite{2010AJ....140..631B} (e.g., the split bright SGB or the double
red MS) and many additional peculiarities are clearly visible in this
$m_{\rm F275W}$ vs. $m_{\rm F275W}-m_{\rm F336W}$ Hess diagram. A
deltailed discussion of the complex population multiplicity in
\wcen\ will be the subject of a forthcoming paper.

\section{Artificial-star tests}
\label{s:comp} 

It is well known that \wcen\ hosts multiple populations of stars that
occupy well-defined and distinct sequences on the CMD at all
evolutionary stages. These sequences typically span a wide range in
color at any given magnitude. For this reason, instead of following
the usual practice of generating a list of artificial stars (ASs)
based on a single fiducial line on the CMD, we populated the
instrumental F606W vs.\ F438W$-$F606W CMD with ASs with colors that,
at any given magnitude level, span the same color range of real stars
(panel (a) of Fig.~\ref{f:as1}).  We simulated two strips of
stars:\ the MS-SGB-RGB strip and the WD-HB strip.  We generated a
total of 550\,000 ASs, and designed to be more abundant towards the
faint limit, in order to better constrain the completeness of faint
stars, according to the following partition:\ 12.5\% in the
instrumental F606W magnitude range $-$13.8 (saturation) to $-$10,
14.5\% between $-$10 and $-$8, 18\% between $-$8 and $-$6.5, 22\%
between $-$6.5 and $-$5.5, and 33\% between $-$5.5 and $-$3. Within
each magnitude interval, ASs have a random F606W and a random
F438W$-$F606W color within the boundaries of the two strips. Then, we
associated to each AS a random X,Y position on the master frame with
the only requirement that an AS must be present in at least one single
exposure in either F606W or F438W. (A random X,Y position is justified
by the fact that the stellar density does not vary significantly
within our FoV, which is comparable in size to the core radius of the
cluster.)

\begin{figure*}[!t]
\centering
\includegraphics[width=\textwidth]{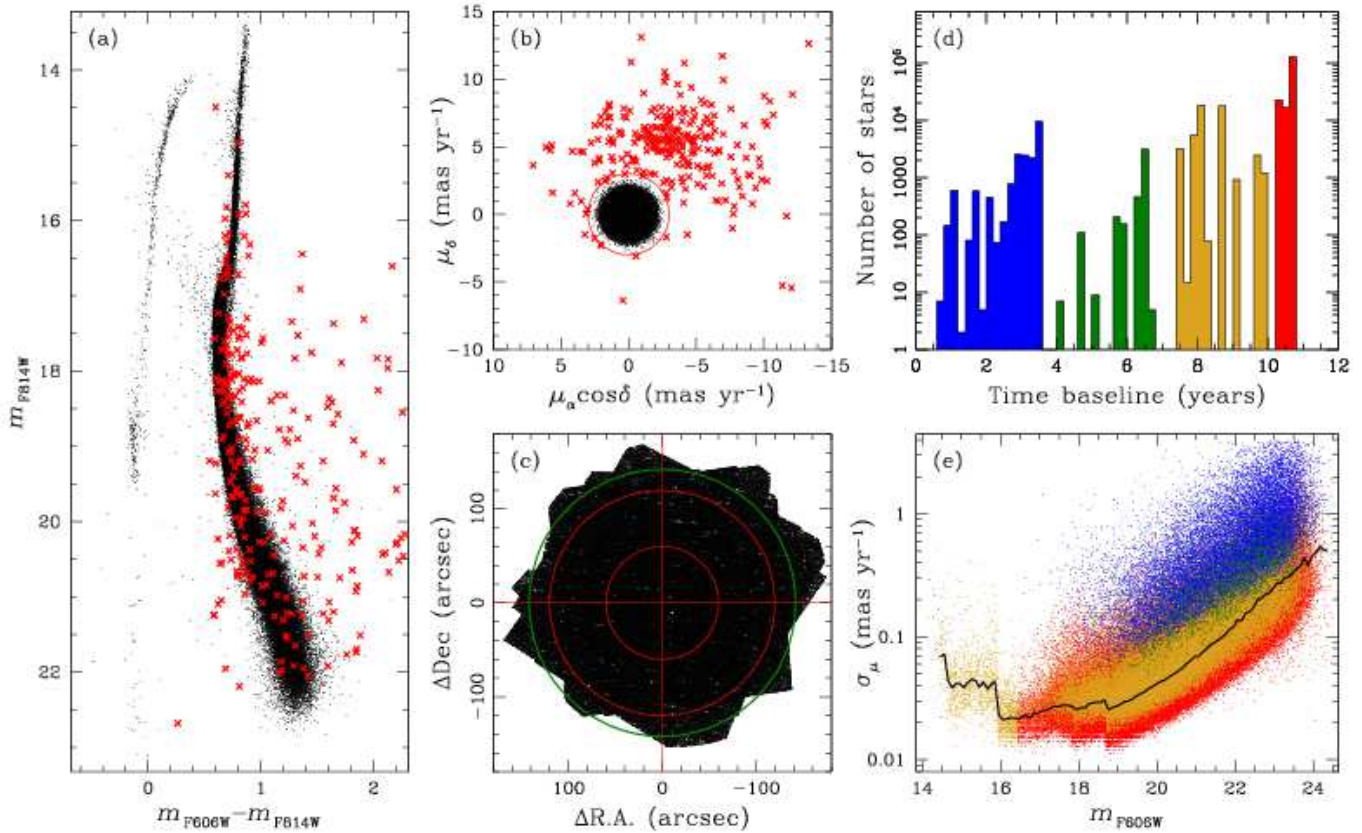}
\caption{\small{(a) $m_{\rm F814W}$ vs. $m_{\rm F606W}-m_{\rm F814W}$
    CMD of stars with photometric RMS $<0.2$ mag and PM error $<0.2$
    mas\,${\rm yr}^{-1}$ (PMs are from
    \citealt{2014ApJ...797..115B}). The vector-point diagram of these
    stars is shown in panel (b). Cluster members (black dots in both
    panels) are arbitrarily separated from field stars (red crosses in
    both panels) by the red circle of radius 3 mas\,yr$^{-1}$. The FoV
    of all stars with a PM measurement is in panel (c). The radius of
    inner and outer red circles is 1 and 2 arcmin, respectively. The
    green circle, at $2\farcm37$ marks the core radius of the
    cluster. (d) histogram of the distribution of the time baselines
    used to compute the PM of each star. We defined 4 groups of stars
    (color-coded from blue to red) having increasingly larger time
    baselines. The PM error as a function of the $m_{\rm F606W}$
    magnitude is in panel (e). Stars are color-coded as in panel
    (d). The 50th percentile of the error distribution is highlighted
    by the black line.}}
\label{f:pm}
\end{figure*}

\begin{table}
\centering
\scriptsize{
\begin{tabular}{ccl}
\multicolumn{3}{c}{\textsc{Table~3}}\\
\multicolumn{3}{c}{\textsc{Column-by-column information contained in the}}\\
\multicolumn{3}{c}{\textsc{astrometric file}}\\
\hline\hline
Col.&Name (unit)&Explanation\\
\hline
1   &ID         &Identification Number\\
2   &X (pix)    & X-position on the master frame\\
3   &Y (pix)    & Y-position on the master frame\\
4   &R.A. (hms) & Right Ascension (J2000)\\
5   &Dec. (deg) & Declination (J2000)\\
6   &$\mu_{\alpha}\cos \delta$ (\masyr)& PM along R.A. \\
7   &$\mu_{\delta}$ (\masyr)& PM along Dec.\\
8   &$\sigma_{\mu_{\alpha}\cos \delta}$ (\masyr)& 1-$\sigma$ uncertainty in $\mu_{\alpha} \cos \delta$ computed using\\
    &                                    & actual residuals\\
9   &$\sigma_{\mu_{\delta}}$ (\masyr)& 1-$\sigma$ uncertainty in $\mu_{\delta}$ computed using actual\\
    &                           &  residuals\\
10  &err$_{\mu_{\alpha} \cos \delta}$ (\masyr)& 1-$\sigma$ uncertainty in $\mu_{\alpha} \cos \delta$ computed using\\
    &                                 & expected errors\\
11  &err$_{\mu_{\delta}}$ (\masyr)& 1-$\sigma$ uncertainty in $\mu_{\delta}$ computed using expected\\
    &                         & errors\\
12  &$\chi^2_{\mu_{\alpha}\cos \delta}$& Reduced $\chi^2$ of the fit of the R.A.-component of\\
    &                          & the motion\\
13  &$\chi^2_{\mu_\delta}$& Reduced $\chi^2$ of the fit of the Dec.-component of\\
    &                 & the  motion\\
14  &$\sigma_{\overline{x}}$ (pix)&1-$\sigma$ uncertainty in the intercept of the PM fit for\\
    &                         & the R.A.-component using actual residuals\\
15  &$\sigma_{\overline{y}}$ (pix)&1-$\sigma$ uncertainty in the intercept of the PM fit for\\
    &                          & the Dec.-component using actual residuals\\
16  & time (yr)& Time baseline, in Julian years\\ 
17  &err$_{\overline{x}}$ (pix)&1-$\sigma$ uncertainty in the intercept of the PM fit for\\
    &                       & the R.A.-component  using expected errors\\
18  &err$_{\overline{y}}$ (pix)&1-$\sigma$ uncertainty in the intercept of the PM fit for\\
    &                      & the Dec.-component  using expected errors\\
19  &U$_{\rm ref}$& Flag: 1=reference star, 0=otherwise\\
20  &N$_{\rm found}$&Total number of data points available for the PM\\
    &            & fits\\
21  &N$_{\rm used}$&Final number of data points used for the PM fits\\
22  &$\Delta \mu_{\alpha}\cos \delta$ (\masyr)& Difference in $\mu_{\alpha}\cos \delta$ between locally-corrected\\
    &                                       &and amplifier-based PMs. Add to column 6 to\\
    &                                       &obtain locally-corrected PMs.\\
23 &$\Delta \mu_{\delta}$ (\masyr) &  Difference in $\mu_{\delta}$ between locally-corrected and\\
   &                             & amplifier-based PMs. Add to column 5 to obtain\\
   &                             & locally-corrected PMs.\\
\hline\hline
\end{tabular}}
\end{table}

\begin{table}
\centering
\scriptsize{
\begin{tabular}{ccl}
\multicolumn{3}{c}{\textsc{Table~4}}\\
\multicolumn{3}{c}{\textsc{Column-by-column information contained in the}}\\
\multicolumn{3}{c}{\textsc{Method-one photometric file}}\\
\hline\hline
Col.&Name (unit)&Explanation\\
\hline
(1) & $m_{\rm filter}$ (mag)& Vega-mag photometry\\
(2) & RMS$_{\rm filter}$ (mag)& Photometric RMS\\
(3) & $N_{\rm f}$& Number of exposures a star is found in\\
(4) & $N_{\rm g}$& Number of good measurements \\
(5) & \texttt{QFIT}&Quality-of-fit parameter\\
(6) & \textit{o}&Fraction of light in the aperture due to neighbors\\
(7) & sky (counts)& Local sky-background value\\
(8) & RMS$_{\rm sky}$ (counts)& Sky RMS\\
(9) & SAT& Flag: 1=saturated, 0=unsaturated\\
\hline\hline
\end{tabular}}
\end{table}

\begin{table}
\centering
\scriptsize{
\begin{tabular}{ccl}
\multicolumn{3}{c}{\textsc{Table~5}}\\
\multicolumn{3}{c}{\textsc{Column-by-column information contained in the}}\\
\multicolumn{3}{c}{\textsc{Method-two and Method-three photometric files}}\\
\hline\hline
Col.&Name (unit)&Explanation\\
\hline
(1) & $m_{\rm filter}$ (mag)& Vega-mag photometry\\
(2) & RMS$_{\rm filter}$ (mag)& Photometric RMS\\
(3) & $N_{\rm f}$& Number of exposures a star is found in\\
(4) & $N_{\rm g}$& Number of good measurements \\
(5) & \texttt{QFIT}&Quality-of-fit parameter\\
(6) & \textit{o}&Fraction of light in the aperture due to neighbors\\
\hline\hline
\end{tabular}}
\end{table}

\begin{table}
\centering
\scriptsize{
\begin{tabular}{ccl}
\multicolumn{3}{c}{\textsc{Table~6}}\\
\multicolumn{3}{c}{\textsc{Column-by-column information contained in the}}\\
\multicolumn{3}{c}{\textsc{Artificial-star-tests file}}\\
\hline\hline
Col.&Name (unit)&Explanation\\
\hline
(1) & X$_{\rm input}$ (pix)& Input X position\\
(2) & Y$_{\rm input}$ (pix)& Input Y position\\
(3) & F606W$_{\rm input}$ (mag)& Input F606W instrumental magnitude\\
(4) & F438W$_{\rm input}$ (mag)& Input F438W instrumental magnitude\\
(5) & X$_{\rm output}$ (pix)& Recovered X position\\
(6) & Y$_{\rm output}$ (pix)& Recovered Y position\\
(7) & F606W$_{\rm output}^{\rm method one}$ (mag)& Recovered F606W magnitude\\
(8) & F438W$_{\rm output}^{\rm method one}$ (mag)& Recovered F438W magnitude\\
(9) & F606W$_{\rm output}^{\rm method two}$ (mag)& Recovered F606W magnitude\\
(10) & F438W$_{\rm output}^{\rm method two}$ (mag)& Recovered F438W magnitude\\
(11) & F606W$_{\rm output}^{\rm method three}$ (mag)& Recovered F606W magnitude\\
(12) & F438W$_{\rm output}^{\rm method three}$ (mag)& Recovered F438W magnitude\\
\hline\hline
\end{tabular}}
\end{table}

\begin{table*}
\centering
\scriptsize{
\begin{tabular}{ccccccccccccc}
\multicolumn{13}{c}{\textsc{Table~7}}\\
\multicolumn{13}{c}{\textsc{Extract of the astrometric file}}\\
\hline\hline
ID&X&Y&R.A.&Dec.&$\mu_{\alpha}\cos \delta$&$\mu_{\delta}$&$\sigma_{\mu_{\alpha}\cos \delta}$&$\sigma_{\mu_{\delta}}$&err$_{\mu_{\alpha}\cos \delta}$&err$_{\mu_{\delta}}$&$\chi^2_{\mu_{\alpha}\cos\delta}$&$\rightarrow$\\
(1)&(2)&(3)&(4)&(5)&(6)&(7)&(8)&(9)&(10)&(11)&(12)&\dots\\
\hline
     1& 3376.59&  353.28& 13:26:35.861& $-$47:26:18.38& $\!\!\!-999.999\!\!\!$& $\!\!\!-999.999\!\!\!$& $\!\!\!-999.999\!\!\!$& $\!\!\!-999.999\!\!\!$& $\!\!\!-999.999\!\!\!$& $\!\!\!-999.999\!\!\!$& $\!\!\!-999.999\!\!\!$&\dots\\
     2& 3389.29&  366.12& 13:26:35.928& $-$47:26:18.61& $\!\!\!-999.999\!\!\!$& $\!\!\!-999.999\!\!\!$& $\!\!\!-999.999\!\!\!$& $\!\!\!-999.999\!\!\!$& $\!\!\!-999.999\!\!\!$& $\!\!\!-999.999\!\!\!$& $\!\!\!-999.999\!\!\!$&\dots\\
     3& 3408.94&  364.30& 13:26:35.994& $-$47:26:18.20& $\!\!\!-999.999\!\!\!$& $\!\!\!-999.999\!\!\!$& $\!\!\!-999.999\!\!\!$& $\!\!\!-999.999\!\!\!$& $\!\!\!-999.999\!\!\!$& $\!\!\!-999.999\!\!\!$& $\!\!\!-999.999\!\!\!$&\dots\\
     4& 3274.62&  364.53& 13:26:35.523& $-$47:26:20.60& $\!\!\!-999.999\!\!\!$& $\!\!\!-999.999\!\!\!$& $\!\!\!-999.999\!\!\!$& $\!\!\!-999.999\!\!\!$& $\!\!\!-999.999\!\!\!$& $\!\!\!-999.999\!\!\!$& $\!\!\!-999.999\!\!\!$&\dots\\
     5& 3275.12&  375.87& 13:26:35.545& $-$47:26:20.99& $\!\!\!-999.999\!\!\!$& $\!\!\!-999.999\!\!\!$& $\!\!\!-999.999\!\!\!$& $\!\!\!-999.999\!\!\!$& $\!\!\!-999.999\!\!\!$& $\!\!\!-999.999\!\!\!$& $\!\!\!-999.999\!\!\!$&\dots\\
     6& 3282.05&  360.71& 13:26:35.543& $-$47:26:20.33& $\!\!\!-999.999\!\!\!$& $\!\!\!-999.999\!\!\!$& $\!\!\!-999.999\!\!\!$& $\!\!\!-999.999\!\!\!$& $\!\!\!-999.999\!\!\!$& $\!\!\!-999.999\!\!\!$& $\!\!\!-999.999\!\!\!$&\dots\\
     7& 3287.16&  355.83& 13:26:35.552& $-$47:26:20.06& $\!\!\!-999.999\!\!\!$& $\!\!\!-999.999\!\!\!$& $\!\!\!-999.999\!\!\!$& $\!\!\!-999.999\!\!\!$& $\!\!\!-999.999\!\!\!$& $\!\!\!-999.999\!\!\!$& $\!\!\!-999.999\!\!\!$&\dots\\
     8& 3292.79&  330.71& 13:26:35.528& $-$47:26:19.07& $\!\!\!-999.999\!\!\!$& $\!\!\!-999.999\!\!\!$& $\!\!\!-999.999\!\!\!$& $\!\!\!-999.999\!\!\!$& $\!\!\!-999.999\!\!\!$& $\!\!\!-999.999\!\!\!$& $\!\!\!-999.999\!\!\!$&\dots\\
     9& 3300.98&  366.09& 13:26:35.619& $-$47:26:20.18& $\!\!\!-999.999\!\!\!$& $\!\!\!-999.999\!\!\!$& $\!\!\!-999.999\!\!\!$& $\!\!\!-999.999\!\!\!$& $\!\!\!-999.999\!\!\!$& $\!\!\!-999.999\!\!\!$& $\!\!\!-999.999\!\!\!$&\dots\\
 \dots&   \dots&   \dots& \dots      & \dots      & \dots    & \dots    & \dots    & \dots    & \dots    & \dots    & \dots    &\dots\\
   194& 3657.93&  543.43& 13:26:37.181& $-$47:26:20.13&    1.3353&    0.2693&    0.0631&    0.1394&    0.0584&    0.0758&    0.7437&\dots\\
   195& 3657.86&  598.49& 13:26:37.277& $-$47:26:22.09&    0.1426& $-$0.6087&    0.0414&    0.0429&    0.0313&    0.0457&    0.3711&\dots\\
   196& 3659.38&  580.75& 13:26:37.251& $-$47:26:21.43& $-$0.2050& $-$0.7416&    0.0403&    0.0293&    0.0447&    0.0331&    0.4803&\dots\\
   197& 3664.87&  587.12& 13:26:37.282& $-$47:26:21.56& $-$1.1008& $-$0.0282&    0.0591&    0.0313&    0.0416&    0.0278&    1.1367&\dots\\
   198& 3666.17&  560.03& 13:26:37.239& $-$47:26:20.57&    0.8398&    1.0971&    0.0314&    0.0538&    0.0312&    0.0490&    0.4734&\dots\\
   199& 3666.51&  598.08& 13:26:37.307& $-$47:26:21.92& $-$0.0023&    0.1116&    0.3434&    0.2926&    0.2674&    0.2653&    1.2947&\dots\\
 \dots&   \dots&   \dots& \dots      & \dots      & \dots    & \dots    & \dots    & \dots    & \dots    & \dots    & \dots    &\dots\\
\hline
&$\rightarrow$&$\chi^2_{\mu_\delta}$&$\sigma_{\overline{x}}$&$\sigma_{\overline{y}}$&time&err$_{\overline{x}}$&err$_{\overline{y}}$&U$_{\rm ref}$&N$_{\rm found}$&N$_{\rm used}$&$\Delta \mu_{\alpha}\cos \delta$&$\Delta \mu_{\delta}$\\
&\dots&(13)&(14)&(15)&(16)&(17)&(18)&(19)&(20)&(21)&(22)&(23)\\
\hline
&\dots& $\!\!\!-999.999\!\!\!$& $\!\!\!-999.999\!\!\!$& $\!\!\!-999.999\!\!\!$& $\!\!\!-999.999\!\!\!$& $\!\!\!-999.999\!\!\!$& $\!\!\!-999.999\!\!\!$&$\!\!-999\!\!$&$\!\!-999\!\!$&$\!\!-999\!\!$& $\!\!\!-999.999\!\!\!$& $\!\!\!-999.999\!\!\!$\\
&\dots& $\!\!\!-999.999\!\!\!$& $\!\!\!-999.999\!\!\!$& $\!\!\!-999.999\!\!\!$& $\!\!\!-999.999\!\!\!$& $\!\!\!-999.999\!\!\!$& $\!\!\!-999.999\!\!\!$&$\!\!-999\!\!$&$\!\!-999\!\!$&$\!\!-999\!\!$& $\!\!\!-999.999\!\!\!$& $\!\!\!-999.999\!\!\!$\\
&\dots& $\!\!\!-999.999\!\!\!$& $\!\!\!-999.999\!\!\!$& $\!\!\!-999.999\!\!\!$& $\!\!\!-999.999\!\!\!$& $\!\!\!-999.999\!\!\!$& $\!\!\!-999.999\!\!\!$&$\!\!-999\!\!$&$\!\!-999\!\!$&$\!\!-999\!\!$& $\!\!\!-999.999\!\!\!$& $\!\!\!-999.999\!\!\!$\\
&\dots& $\!\!\!-999.999\!\!\!$& $\!\!\!-999.999\!\!\!$& $\!\!\!-999.999\!\!\!$& $\!\!\!-999.999\!\!\!$& $\!\!\!-999.999\!\!\!$& $\!\!\!-999.999\!\!\!$&$\!\!-999\!\!$&$\!\!-999\!\!$&$\!\!-999\!\!$& $\!\!\!-999.999\!\!\!$& $\!\!\!-999.999\!\!\!$\\
&\dots& $\!\!\!-999.999\!\!\!$& $\!\!\!-999.999\!\!\!$& $\!\!\!-999.999\!\!\!$& $\!\!\!-999.999\!\!\!$& $\!\!\!-999.999\!\!\!$& $\!\!\!-999.999\!\!\!$&$\!\!-999\!\!$&$\!\!-999\!\!$&$\!\!-999\!\!$& $\!\!\!-999.999\!\!\!$& $\!\!\!-999.999\!\!\!$\\
&\dots& $\!\!\!-999.999\!\!\!$& $\!\!\!-999.999\!\!\!$& $\!\!\!-999.999\!\!\!$& $\!\!\!-999.999\!\!\!$& $\!\!\!-999.999\!\!\!$& $\!\!\!-999.999\!\!\!$&$\!\!-999\!\!$&$\!\!-999\!\!$&$\!\!-999\!\!$& $\!\!\!-999.999\!\!\!$& $\!\!\!-999.999\!\!\!$\\
&\dots& $\!\!\!-999.999\!\!\!$& $\!\!\!-999.999\!\!\!$& $\!\!\!-999.999\!\!\!$& $\!\!\!-999.999\!\!\!$& $\!\!\!-999.999\!\!\!$& $\!\!\!-999.999\!\!\!$&$\!\!-999\!\!$&$\!\!-999\!\!$&$\!\!-999\!\!$& $\!\!\!-999.999\!\!\!$& $\!\!\!-999.999\!\!\!$\\
&\dots& $\!\!\!-999.999\!\!\!$& $\!\!\!-999.999\!\!\!$& $\!\!\!-999.999\!\!\!$& $\!\!\!-999.999\!\!\!$& $\!\!\!-999.999\!\!\!$& $\!\!\!-999.999\!\!\!$&$\!\!-999\!\!$&$\!\!-999\!\!$&$\!\!-999\!\!$& $\!\!\!-999.999\!\!\!$& $\!\!\!-999.999\!\!\!$\\
&\dots& $\!\!\!-999.999\!\!\!$& $\!\!\!-999.999\!\!\!$& $\!\!\!-999.999\!\!\!$& $\!\!\!-999.999\!\!\!$& $\!\!\!-999.999\!\!\!$& $\!\!\!-999.999\!\!\!$&$\!\!-999\!\!$&$\!\!-999\!\!$&$\!\!-999\!\!$& $\!\!\!-999.999\!\!\!$& $\!\!\!-999.999\!\!\!$\\
&\dots& \dots    & \dots    & \dots    & \dots     & \dots    & \dots    &\dots&\dots&\dots&\dots     &\dots     \\
&\dots&    1.2052&    0.0070&    0.0157&    8.73723&    0.0065&    0.0084&    0&   17&   16& $-$0.0745& $-$0.1032\\
&\dots&    0.7171&    0.0046&    0.0048&    8.73723&    0.0035&    0.0051&    1&   17&   17& $-$0.0455& $-$0.0673\\
&\dots&    0.3010&    0.0045&    0.0031&    8.73723&    0.0050&    0.0037&    0&   17&   14& $-$0.0684& $-$0.0624\\
&\dots&    0.5465&    0.0066&    0.0035&    8.73723&    0.0046&    0.0031&    1&   17&   17&    0.0032&    0.1158\\
&\dots&    1.0716&    0.0035&    0.0060&    8.73723&    0.0035&    0.0054&    1&   17&   17& $-$0.0128&    0.0851\\
&\dots&    1.2767&    0.0381&    0.0325&    8.73722&    0.0299&    0.0297&    0&   12&   12&    0.1921& $-$0.1553\\
&\dots& \dots    & \dots    & \dots    & \dots     & \dots    & \dots    &\dots&\dots&\dots&\dots     &\dots     \\
\hline\hline
&&&&&&&&&&&&\\
\end{tabular}}
\end{table*}

\begin{table}
\centering
\scriptsize{
\begin{tabular}{ccccccccc}
\multicolumn{9}{c}{\textsc{Table~8}}\\
\multicolumn{9}{c}{\textsc{Extract of the method-one F606W photometric file}}\\
\hline\hline
$m_{\rm F606W}$&RMS$_{\rm F606W}$&$N_{\rm f}$&$N_{\rm g}$&\texttt{QFIT}&\textit{o}&sky&RMS$_{\rm sky}$&SAT\\
(1)&(2)&(3)&(4)&(5)&(6)&(7)&(8)&(9)\\
\hline
   23.213&   $-$99.999&   1 &  1&  0.969 &      0.20 &      0.0 &      5.6 &  0\\
   18.107&   $-$99.999&   1 &  1&  0.999 &      0.00 &    149.8 &    118.0 &  0\\
   20.311&   $-$99.999&   1 &  1&  0.999 &      0.00 &     11.7 &     18.7 &  0\\
   18.140&   $-$99.999&   1 &  1&  1.000 &      0.01 &    217.3 &    130.0 &  0\\
   19.908&   $-$99.999&   1 &  1&  0.999 &      0.01 &     36.0 &     25.4 &  0\\
   17.678&   $-$99.999&   1 &  1&  1.000 &      0.00 &    291.6 &    196.2 &  0\\
   21.026&   $-$99.999&   1 &  1&  0.995 &      0.17 &     39.5 &     39.9 &  0\\
   19.283&   $-$99.999&   1 &  1&  0.999 &      0.00 &     45.4 &     44.1 &  0\\
   21.675&   $-$99.999&   1 &  1&  0.997 &      0.01 &      0.0 &      6.8 &  0\\
   19.286&   $-$99.999&   1 &  1&  0.999 &      0.01 &     88.2 &     78.1 &  0\\
\dots&\dots&\dots&\dots&\dots&\dots&\dots&\dots&\dots\\
\hline\hline
\end{tabular}}
\end{table}
\begin{table}
\centering
\scriptsize{
\begin{tabular}{cccccc}
\multicolumn{6}{c}{\textsc{Table~9}}\\
\multicolumn{6}{c}{\textsc{Extract of the method-two F606W photometric file}}\\
\hline\hline
$m_{\rm F606W}$&RMS$_{\rm F606W}$&$N_{\rm f}$&$N_{\rm g}$&\texttt{QFIT}&\textit{o}\\
(1)&(2)&(3)&(4)&(5)&(6)\\
\hline
   23.114&   $-99.999$ &  1 &  1 & 0.953   &    0.04\\
   18.133&   $-99.999$ &  1 &  1 & 0.999   &    0.00\\
   20.337&   $-99.999$ &  1 &  1 & 0.999   &    0.00\\
   18.132&   $-99.999$ &  1 &  1 & 1.000   &    0.00\\
   19.914&   $-99.999$ &  1 &  1 & 0.999   &    0.00\\
   17.680&   $-99.999$ &  1 &  1 & 1.000   &    0.00\\
   21.083&   $-99.999$ &  1 &  1 & 0.995   &    0.03\\
   19.274&   $-99.999$ &  1 &  1 & 0.999   &    0.00\\
   21.681&   $-99.999$ &  1 &  1 & 0.997   &    0.00\\
   19.307&   $-99.999$ &  1 &  1 & 1.000   &    0.00\\
\dots&\dots&\dots&\dots&\dots&\dots\\
\hline\hline
\end{tabular}}
\end{table}

\begin{table}
\centering
\scriptsize{
\begin{tabular}{cccccc}
\multicolumn{6}{c}{\textsc{Table~10}}\\
\multicolumn{6}{c}{\textsc{Extract of the method-three F606W photometric file}}\\
\hline\hline
$m_{\rm F606W}$&RMS$_{\rm F606W}$&$N_{\rm f}$&$N_{\rm g}$&\texttt{QFIT}&\textit{o}\\
(1)&(2)&(3)&(4)&(5)&(6)\\
\hline
   23.026 &  $-99.999$ &  1  & 1&  0.987  &     0.03\\
   18.151 &  $-99.999$ &  1  & 1&  1.000  &     0.00\\
   20.359 &  $-99.999$ &  1  & 1&  1.000  &     0.00\\
   18.128 &  $-99.999$ &  1  & 1&  1.000  &     0.00\\
   19.926 &  $-99.999$ &  1  & 1&  1.000  &     0.00\\
   17.679 &  $-99.999$ &  1  & 1&  1.000  &     0.00\\
   21.121 &  $-99.999$ &  1  & 1&  0.998  &     0.02\\
   19.265 &  $-99.999$ &  1  & 1&  1.000  &     0.00\\
   21.683 &  $-99.999$ &  1  & 1&  1.000  &     0.00\\
   19.322 &  $-99.999$ &  1  & 1&  1.000  &     0.00\\
\dots&\dots&\dots&\dots&\dots&\dots\\
\hline\hline
\end{tabular}}
\end{table}

\begin{sidewaystable}[!h]
\centering
\begin{tabular}{cccccccccccc}
\multicolumn{12}{c}{\textsc{Table~11}}\\
\multicolumn{12}{c}{\textsc{Extract of the Artificial-star-tests file}}\\
\hline\hline
X$_{\rm input}$&Y$_{\rm input}$&F606W$_{\rm input}$&F438W$_{\rm input}$&X$_{\rm output}$&Y$_{\rm output}$
&F606W$_{\rm output}^{\rm method one}$&F438W$_{\rm output}^{\rm method one}$&F606W$_{\rm output}^{\rm method two}$
&F438W$_{\rm output}^{\rm method two}$&F606W$_{\rm output}^{\rm method three}$&F438W$_{\rm output}^{\rm method three}$\\
(1)&(2)&(3)&(4)&(5)&(6)&(7)&(8)&(9)&(10)&(11)&(12)\\
\hline
 3314.8723&  297.4116&  $-$3.7516&  $-$2.9688& 3240.50&  240.50&  $-$0.0000&  $-$0.0000&  $-$0.0000&  $-$0.0000&  $-$0.0000&  $-$0.0000\\
 3329.1575&  307.4878& $-$11.5694& $-$13.1609& 3329.14&  307.48& $-$11.5779& $-$13.1602& $-$11.5694& $-$13.1614& $-$11.5634& $-$13.1580\\
 3321.9863&  307.7570&  $-$3.3748&  $-$1.7733& 3326.48&  304.04&  $-$7.2610&  $-$6.7243&  $-$7.2887&  $-$6.7171&  $-$7.3347&  $-$6.8063\\
 3300.7180&  308.4292&  $-$3.6456&  $-$4.1334& 3304.18&  312.34& $-$10.5523& $-$11.0767& $-$10.5308& $-$11.0618& $-$10.5156& $-$11.0500\\
 3323.2595&  311.8160&  $-$4.9417&  $-$5.8025& 3326.48&  304.04&  $-$7.2663&  $-$6.7113&  $-$7.2907&  $-$6.7132&  $-$7.3340&  $-$6.8051\\
 3323.9541&  313.1805&  $-$5.1536&  $-$6.1770& 3326.48&  304.04&  $-$7.2722&  $-$6.7177&  $-$7.2923&  $-$6.7143&  $-$7.3346&  $-$6.8036\\
 3319.8965&  314.8643&  $-$7.1340&  $-$8.5612& 3319.90&  314.88&  $-$7.1519&  $-$8.5898&  $-$7.1488&  $-$8.6031&  $-$7.1157&  $-$8.5994\\
 3316.6787&  315.6735&  $-$5.2196&  $-$6.5196& 3320.06&  322.27&  $-$9.5062&  $-$9.8308&  $-$9.5134&  $-$9.8132&  $-$9.5143&  $-$9.8232\\
 3324.8562&  317.0846&  $-$7.7829&  $-$9.1057& 3324.92&  317.07&  $-$7.8698&  $-$9.0958&  $-$7.7984&  $-$9.1173&  $-$7.7987&  $-$9.1244\\
 3296.6982&  319.2882&  $-$5.6162&  $-$4.4644& 3240.50&  240.50&  $-$0.0000&  $-$0.0000&  $-$0.0000&  $-$0.0000&  $-$0.0000&  $-$0.0000\\
\dots&\dots&\dots&\dots&\dots&\dots&\dots&\dots&\dots&\dots&\dots&\dots\\
\hline\hline
\end{tabular}
\end{sidewaystable}

ASs are then added, measured and removed one at a time by KS2, using
all three photometric methods.  This way, ASs never interfere with
each other. Each AS test thus consists of a set of four input
parameters (X,Y positions and instrumental magnitudes in F606W and
F438W) and the same output parameters for the nearest found star for
the three methods.  The user can then determine whether the recovered
star corresponds to the inserted star. Typically, if the input and
output positions agree to within 0.5 pixel and the fluxes agree to
within 0.75 mag, then a star can be considered found, but different
scientific investigations might require different thresholds.

Panel (b) of Fig.~\ref{f:as1} shows a color-coded Hess diagram of the
output CMD based on method-two photometry. The 50\% completeness level
is reached at about F606W$\sim$$-$6.33 along the MS, and about
F606W$\sim$$-$5.95 along the WD cooling sequence ($m_{\rm
  F606W}$$\sim$23.51 and $\sim$23.89, respectively).

Another way to analyze the completeness level of the catalog is to
look at its spatial distribution (panel c of Fig.~\ref{f:as1}). We
considered only the input stars with F606W between instrumental
magnitude $-$7 and $-$5 (i.e., around the 50\% completeness level). In
general, the completeness distribution around the center of the FoV is
flat, as we would expect given the flat stellar density profile of the
cluster in this region. On the other hand, the outskirts of our FoV
are affected by a much lower completeness level. These regions are
typically mapped only by one or two single exposures, while the
mapping close to the center of the FoV is provided by more 60
exposures. As a consequence, we were unable to recover as many stars
in the outskirts of the FoV as in the center.  Another feature visible
in panel (c) is that, despite the fairly constant star density in our
FoV, there are patchy areas where the local completeness level drops
sizably. This happens in the vicinity of saturated stars, due to being
masked during the finding phase.

As we already discussed in Section~\ref{ss:spp}, KS2 provides three
different methods of measuring stellar photometry. Each of these
methods works best in different magnitude regimes. This can also be
seen through the comparison of the completeness level reached by each
method at any given magnitude. As an example, for a more homogeneous
comparison, we selected all ASs within 3000 pixels from the center of
the master frame (black circle in panel c), where we are not affected
by mapping biases. The completeness level of the three photometric
methods in this region is shown in Fig.~\ref{f:as2}. We color-coded
method one in red, method two in green, and method three in blue. The
inset shows a blown-up region around the 50\% completeness level. The
magnitudes at which each method reaches 50\% completeness are marked
for clarity.  The reason method one is less complete at the faint end
is because these stars are less likely to generate a distinct and
isolated peak in a single exposure. The other methods are not affected
by this limitation and share a very similar completeness curve. Method
three is marginally more complete than method two at the faint end
because it provides slightly more accurate photometric estimates for
the faintest stars.

\section{Astrometry}
\label{s:astro}

We cross-identified stars in our catalog with the stars in the Gaia
data release 1 (Gaia DR1,
\citealt{2016arXiv160904303L})\footnote[16]{http://gea.esac.esa.int/archive/}
within 5 arcmin from the cluster center. Gaia DR1 R.A. and
Dec. positions are given at the reference epoch 2015.0 and at the
equinox J2000, with respect to the International Celestial Reference
System (ICRS).

We found about 700 sources in common, which were used to register our
positions to the Gaia-DR1 absolute astrometric system.  Absolute
stellar positions in our catalog have an accuracy of about 5 mas.

\section{Proper motions}
\label{s:pm} 

In \cite{2014ApJ...797..115B}, we published high-precision PM catalogs
for 22 GCs, including \wcen. The \wcen\ PM catalog is based on several
of the UVIS wide-filter exposures reduced in this work, plus ACS/WFC
exposures taken in 2002 (GO-9442, PI: Cool), 2004 (GO-10252,
PI:\ Anderson) and 2006 (GO-10775, PI:\ Sarajedini), see Table~12 of
\cite{2014ApJ...797..115B} for the complete list of
observations.\footnote[17]{Note that observations taken with filters
  F225W and F275W are not suitable for high-precision PM measurements,
  because of color-dependent residuals present in the UVIS GD
  corrections (\citealt{2011PASP..123..622B}). Moreover, the 2004
  ACS/WFC observations do not overlap with our photometric catalog.}
In the following, we describe how PMs were measured in
\cite{2014ApJ...797..115B} and how we cross-matched their PM catalog
with the photometric catalog presented here.

As described in \cite{2014ApJ...797..115B}, PMs were measured using
the central-overlap method (e.g., \citealt{1971PMcCO..16..267E}), in
which each exposure counts as a stand-alone epoch. Stellar positions
of each exposure are transformed into a common, astrometrically-flat
master frame by means of general six-parameter linear
transformations. We adopted a semi-local transformation approach
where, in each exposure, stars associated with different detector
amplifiers are treated separately. For a given star, all X and Y
transformed positions on the master frame as a function of the epoch
are then least-square fitted with a straight line, the slope of which
is a direct measurement of the stellar motion.  PM errors in the
catalog are measured in two ways:\ (1) using the estimated errors
(based on our knowledge of the typical expected astrometric error of
stars at any given instrumental magnitude) as weights, and (2) using
the actual residuals of the data points around the fitted lines.

Systematic errors in the PMs are mitigated to the extent possible, and
local PM corrections are included in the catalog. We refer the reader
to \cite{2014ApJ...797..115B} for a detailed description of the
reduction method and for an exhaustive discussion about how to
properly reject low-quality PM measurements.  The \wcen\ PM catalog of
\cite{2014ApJ...797..115B} contains a total of 313\,286 sources, and
is complemented with various quality parameters that help to in
identify and remove low-quality measurements (see the detailed
discussion in Sect.~7 of \citealt{2014ApJ...797..115B}).

We cross-identified our photometric catalog with the PM catalog of
\cite{2014ApJ...797..115B}, and found 270\,909 objects in common. The
majority of stars in our photometric catalog without PM measurements
are faint MS stars. In addition, saturated stars also have no PM
measurements. On the other hand, stars that are present in the
\cite{2014ApJ...797..115B} PM catalog but have no counterpart in the
photometric catalog all lie outside the FoV of the photometric
catalog. As we stated at the beginning of this Section, the 2004
ACS/WFC observations do not overlap with our photometric catalog.

Panel (a) of Fig.~\ref{f:pm} shows the $m_{\rm F814W}$ vs. $m_{\rm
  F606W}-m_{\rm F814W}$ CMD for stars that have photometric RMS $<0.2$
mag and PM errors\footnote[18]{The PM error is defined as the sum in
  quadrature of the errors along R.A. and Dec.}  $<0.2$ mas\,yr$^{-1}$
($\sim$5 km$\,$s$^{-1}$ at the distance of \wcen\ of 5.2 kpc,
\citealt{1996AJ....112.1487H}, 2010 edition). The vector-point diagram
of these stars is in panel (b). The vast majority of stars are cluster
members and clump at (0,0), as one would expect since PMs are measured
relative to the bulk motion of the cluster. A much broader, less
populated clump of objects, around ($-4$,6), represents field
stars. On the vector-point diagram we can consider as bona-fide
cluster members (black points in panels (a) and (b)) those stars for
which the motion is within 3 mas\,yr$^{-1}$ (red circle) of the bulk
motion of the cluster . Field stars (outside the circle) are
highlighted with red crosses in both panels (a) and (b).

The FoV of stars with PM measurements is shown in panel (c) of
Fig.~\ref{f:pm}. Units are arcsecs with respect to the cluster's
center (R.A.,Dec.=13:26:47.24,$-$47:28:46.45,
\citealt{2010ApJ...710.1032A}). We drew two red circles of 1 and 2
arcmin for clarity. The green circle in the panel, at $2\farcm37$,
marks the cluster's core radius (\citealt{1996AJ....112.1487H}, 2010
edition).

The histogram of time baseline used to compute the PM of each star is
shown in panel (d) (note the logarithmic scale on the Y axis). For
convenience, we divided the sample into 4 groups with different time
baselines:\ 0--3.8 years (blue), 3.8--7 years (green), 7--10 years
(yellow), and 10--10.6 years (red). The PM of the vast majority of
stars was computed over time baselines larger than 10 yrs (171,276
stars to be precise), while the remaining groups account for 19,726,
4,119, and 50,322, from blue to yellow, respectively. (Note that here
we are considering only stars with photometric RMS $<$0.2 mag and PM
errors $<$0.2 mas$\,$yr$^{-1}$, so that the total number of stars in
the panel do not add up to 270\,909.)

The PM error $\sigma_\mu$ (not to be confused with the velocity
dispersion) of each of these four groups is shown in panel (e), as a
function of the $m_{\rm F606W}$ magnitude.  Clearly, PMs based on
larger time baselines have smaller errors.  The black line is a
running median of the PM errors as a function of
magnitude. Well-measured stars (between $m_{\rm F606W}$=16 and 19)
have a typical PM error of about 25 $\mu$as yr$^{-1}$, or about 0.6 km
s$^{-1}$ at a distance of 5.2 kpc, but they can be as low as 15
$\mu$as yr$^{-1}$ in some cases.

\section{The catalog}
\label{s:cat}

The final catalog is split into several files. There is a single,
general astrometric file (\texttt{ID\_XY\_RD\_PM.dat}), containing
information about stellar positions, PMs, and PM diagnostics
quantities, and a single artificial-star-test file
(\texttt{AS\_IO.dat}).  Then, for each filter, we provide a distinct
file for each photometric method containing magnitudes and quality
parameters (e.g., \texttt{F606W.m1.dat}, \texttt{F606W.m2.dat}, or
\texttt{F606W.m3.dat} for methods one, two and three, respectively).
Each of these files starts with a header containing a column-by-column
description of the contained data, followed the same number of ordered
data lines, one for each star.

The astrometric file contains 49 lines of header information. It then
contains one line for each star, with 23 columns with stellar
positions in both X,Y and R.A.,Dec. units, followed by PM information
coming directly from the PM catalog of \cite{2014ApJ...797..115B} (see
Table~3). If a star has no PM measurements, a flag value of $-999.999$
is adopted for all PM-related columns except for $U_{\rm ref}$,
N$_{\rm found}$ and N$_{\rm used}$, which are instead flagged to
$-999$.

The first six columns of the three photometric files contain Vega-mag
magnitudes and quality parameters for each measured star. In addition,
the method-one files also contain information about the local sky
background, as well as a flag to distinguish between unsaturated and
saturated stars (for which photometry comes from the first-pass
reduction). As a result, the method-one photometric file has a 12-line
header, while method-two and three files have an 8-line header.  If a
star is found in only one exposure, it is not possible to compute its
photometric RMS, and a flag value of $-99.999$ is used. If a star is
not measured in one particular filter, both its magnitude and RMS will
be flagged at $-99.999$.  Tables~4 and 5 list the column-by-column
information of the photometric files.

The \texttt{AS\_IO.dat} AS-test file has 12 lines of header
information, and contains 12 columns for each AS star:\ X$_{\rm
  input}$, Y$_{\rm input}$, F606W$_{\rm input}$ and F438W$_{\rm
  input}$ input values, followed by the same quantities as recovered
by KS2 for the three photometric methods (see Table~6). Input and
output magnitudes are in instrumental units. To convert them into
VEGA-MAG units, the proper $\Delta$mag and ZP(VEGA) values listed in
Table~2 must be added.

Tables~7 to 10 show an extract of the astrometic file and the first
ten lines of the three photometric files for the F606W filter,
respectively.

Together with the catalog, we release to the astronomical community
our 8500$\times$9000 pixels image stacks, in FITS format (one per
filter).  Please note that these image stacks are not meant to be used
to extract high-precision photometry, because both the shape and the
flux of the stars in the stacks are not represented with a high level
of fidelity. The primary use of the image stacks should be that of
finding charts.

\section{Summary}
\label{s:summ} 

We have constructed the most-comprehensive catalog of photometry and
proper motions ever assembled for the core of the GC \wcen\ (or any
other cluster). The catalog, containing over 470$\,$000 stars, is
based on over 650 exposures taken with the WFC3's UVIS and IR channels
in 26 distinct filters, continuously spanning from F225W in the UV to
F160W in the infrared. The photometry is measured in 3 different ways,
each optimized for a different brightness and crowding regime. We
supplemented the photometric catalog with extensive artificial-star
tests.

We cross-identified the photometric catalog with the high-precision PM
compilation of \cite{2014ApJ...797..115B}, and found about 270$\,$000
objects in common. The best-measured stars in the catalog have typical
PM error of $\sim$25 $\mu$as$\,$yr$^{-1}$, or $\sim$0.6
km$\,$s$^{-1}$.

In this first paper of the series, we have described in detail each
stage of the data-reduction processes, and we are making the
astro-photometric catalog, artificial-star tests and the high-quality
image stacks available to the astronomical community. Future papers of
the series will be focused on the analysis and interpretation of the
catalog, in particular on the identification and characterization of
multiple-stellar populations and exotic objects both from the
photometric and astrometric point of view. Paper~II in this series
will be focused on deriving a high-precision differential-reddening
map of the field, and in paper~III we will analyze in great detail the
MPs along the MS of the cluster.

\acknowledgments \noindent \textbf{Acknowledgments.} AB acknowledges
support from STScI grants AR-12656 and AR-12845.  GP acknowledges
partial support by PRIN-INAF 2014 e by the "Progetto di Ateneo 2014
CPDA141214 by Universit\`a di Padova.

\small{
}

\end{document}